\documentclass[11pt,a4paper]{article}
\usepackage[margin=1in]{geometry}

\usepackage{xspace}
\usepackage{graphicx}
\usepackage{array}
\usepackage{booktabs}
\usepackage{subcaption}
\usepackage{float}
\usepackage{gensymb}
\usepackage{clrscode3e}

\newcommand{\visible}{\texttt{visible}\xspace}
\newcommand{\invisible}{\texttt{invisible}\xspace}

\newcommand{\visdac}{\textsc{vis-dac}\xspace}
\newcommand{\visiter}{\textsc{vis-iter}\xspace}
\newcommand{\rrr}{\texttt{R3}\xspace}
\newcommand{\lores}{{\sc LoRes}\xspace}
\newcommand{\hires}{{\sc HiRes}\xspace}
\newcommand{\blockit}{{\sc BlockGrid}\xspace}
\newcommand{\multires}{\textsc{MultiRes}\xspace}
\newcommand{\set}{\texttt{Set1}\xspace}
\newcommand{\bruns}{\texttt{Brunswick}\xspace}
\newcommand{\kaweah}{\texttt{Kaweah}\xspace}
\newcommand{\southport}{\texttt{Southport\xspace}}
\newcommand{\sierra}{\texttt{Sierra\xspace}}
\newcommand{\oahu}{\texttt{Oahu\xspace}}
\newcommand{\app}{\texttt{Appalachians\xspace}}
\newcommand{\clearcr}{\texttt{Clear Creek}\xspace}
\newcommand{\portland}{\texttt{Portland}\xspace}
\newcommand{\wash}{\texttt{Washington}\xspace}

\newtheorem{lemma}{Lemma}

\newtheorem{theorem}{Theorem}

 \def\BibTeX{{\rm B\kern-.05em{\sc i\kern-.025em b}\kern-.08emT\kern-.1667em\lower.7ex\hbox{E}\kern-.125emX}}

\begin{document}

%
\title{Best of Two Worlds: Combing High and Low Resolution for
  Computing Viewsheds on Terrains\footnote{A short version 
  of this paper appeared in the Proceedings of ACM SIGSPATIAL GIS, 2018}}

%
%


\author{
Laura Toma\\Bowdoin College, Maine, USA, email: ltoma@bowdoin.edu
}

\date{}

\begin{abstract}

 The viewshed of a point $v$ on a grid terrain $T$, viewshed$_T(v)$,
  is defined as the set of grid points in $T$ that are visible from $v$.  We
  describe a novel algorithm for computing viewshed$_T(v)$ using a
  multiresolution approach: Given a parameter $k >1$ that represents
  the block size, we create a grid $T'$ which is a lower-resolution
  version of $T$, such that each point in $T'$ corresponds to a block
  of $\lceil \sqrt k \rceil $-by-$\lceil \sqrt k \rceil$ points in
  $T$; $T'$ has size $\Theta(n/k)$, where $n$ is the size of the
  original grid.  The key of our approach is using $T'$ to speed up
  the computation of viewshed$_T(v)$ while not introducing
  approximation.  We compute viewshed$_T(v)$ in two steps: First we
  compute the viewshed of $v$ on $T'$, while maintaining the invariant
  that any block in $T'$ that is labeled as invisible may not contain
  any visible points.  Thus, the first step's role is to use $T'$ to
  filter out blocks in $T$ that are guaranteed to be invisible.
  The second step considers the blocks that were labeled as visible in
  $T'$ and computes the visibility of their points with full accuracy
  using the data in $T$.  Overall the algorithm runs in $O(n + \frac nk
  \lg \frac nk + k \lg k +  l \cdot \lg n)$, where $l$ is the total size of
  visible blocks in $T'$.  When $k = \Omega(1)$ and $l = o(n) $, the running
  time of our algorithm improves on the previous best bound of $O(n
  \lg n)$~\cite{ht:visLI,kreveld:viewshed}.
  We describe a suite of experimental results showing the performance
  of the new  algorithm in practice and a speedup of more than an order of
  magnitude compared to previous algorithms.

\end{abstract}
\maketitle

\section{Introduction}

In recent years digital elevation data available from Lidar technology
at very fine horizontal resolution and vertical accuracy has expanded
the interest in digital terrain modeling and applications. One
fundamental computation on terrains is visibility: two points $a$ and
$b$ on a terrain are considered visible to each other if the interior
of their \emph{line-of-sight} $ab$ (the line segment between them)
lies entirely above the terrain. The most studied visibility problem
on terrains is the \emph{viewshed}: Given a terrain and an arbitrary
viewpoint $v$, the viewshed of $v$ is defined as the set of all points
in the terrain that are visible from $v$. Visibility on terrains has
many applications, including placement of towers in telecommunications,
strategic military planning, archaeology and digital
humanities~\cite{viewshed:greek}; see for
e.g.~\cite{floriani:vissurvey} and the references therein.

In practice terrain data comes as a set of elevation samples, where
each point represents the latitude, longitude, and the measured
elevation of the terrain at that point. A digital terrain model (in
short: a digital terrain) consists of a set of elevation samples and
an interpolation method that describes how to model the elevation of
the terrain at points other than the sample points. The most commonly
used digital terrain is the grid, in which the elevation is sampled
with a uniform grid. A grid terrain is stored as a matrix of elevation
values and is most commonly used with linear, bilinear or
nearest-neighbor interpolation. A different terrain model is the
triangulated irregular network (TIN), see for e.g. De Floriani and
Magillo~\cite{floriani:vissurvey} for a discussion.


Given a digital terrain $T$, in order to determine if a point $p$ is
visible from $v$, one needs to interpolate the elevation of $T$ along
the projection of the line-of-sight $vp$ on the horizontal plane, and
check if the interpolated elevations are below $vp$. Algorithms for
computing visibility differ in what points they chose to interpolate
along the line-of-sight and what interpolation method they use, and thus the
choice of the interpolation method/terrain model immediately impacts
the output and  the running time complexity.

To be useful in practice, viewshed algorithms need to be efficient and
accurate, and these issues have become more important with the
availability of very large, high-resolution data.  The accuracy of the
viewshed needs to be considered in the context of the accuracy of the
data: Consider for e.g. a grid terrain at 100m resolution, which means
that every 100m by 100m cell is represented by one average elevation;
the resolution of the data limits the accuracy of the viewshed, and
small differences in the algorithm or interpolation will likely have
small effects to the overall accuracy. However, terrain models based
on Lidar data can have 1m horizontal resolution and 15cm vertical
accuracy~\cite{lidar101}, which makes it important that viewshed
algorithms do not introduce unnecessary approximation. Grid terrains
at 1m resolution are now widely used and consist of hundreds of
million of points for geographical areas that are relatively small;
for e.g. a grid for Lincoln County, Maine, USA, at 2m resolution, has
$919$ million points. Computing visibility on high-resolution data has
the potential for great accuracy, but require accurate and efficient
algorithms.

\begin{figure*}[t]
  \begin{center} 
    \begin{subfigure}{.24\textwidth} 
      \includegraphics[width=.9\linewidth]{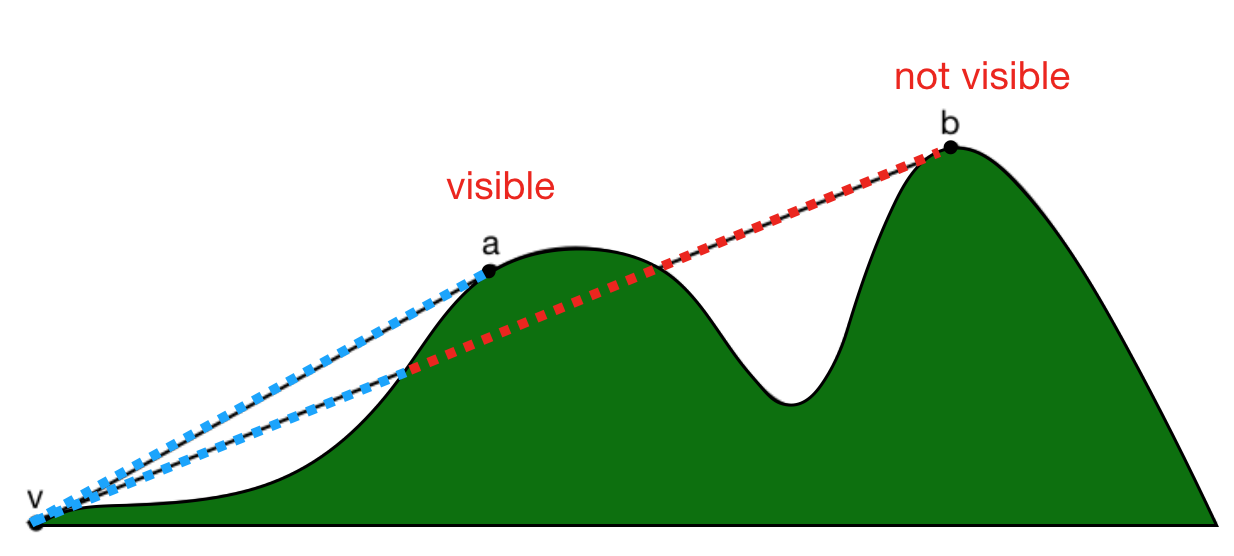}
      \caption{}
    \end{subfigure} 
 \begin{subfigure}{.24\textwidth} 
      \includegraphics[width=.9\linewidth]{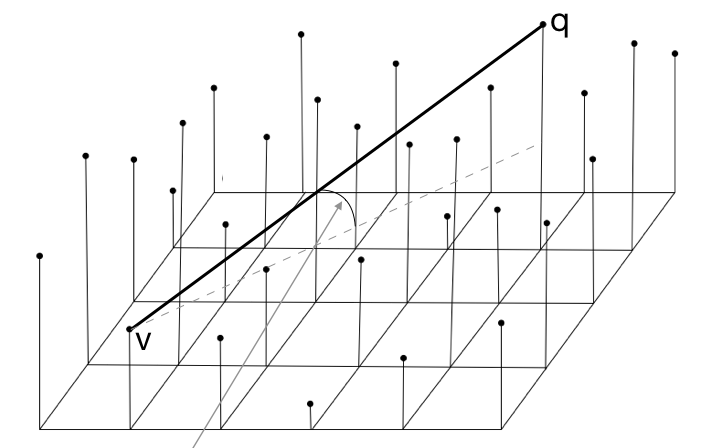}
      \caption{}
    \end{subfigure} 
 \begin{subfigure}{.24\textwidth} 
      \includegraphics[width=.7\linewidth]{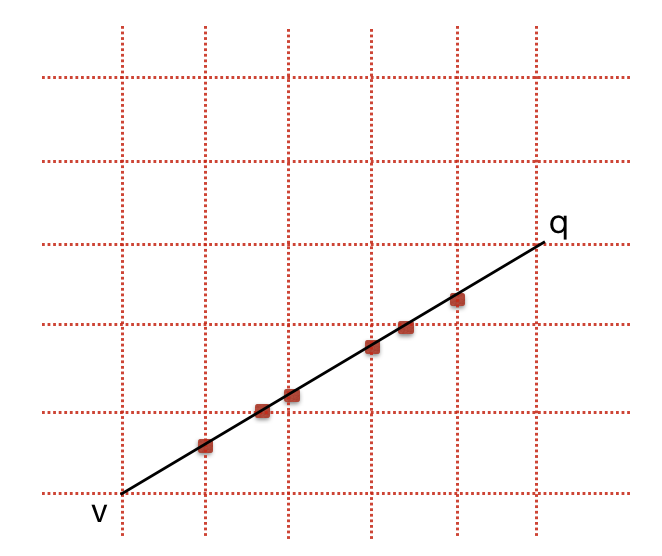}
      \caption{}
 \end{subfigure} 
 \begin{subfigure}{.24\textwidth} 
      \includegraphics[width=.9\linewidth]{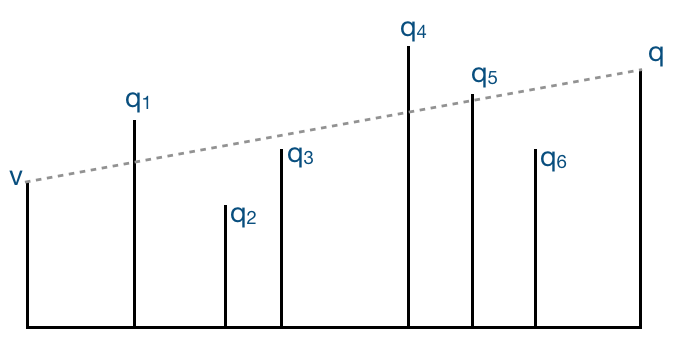}
      \caption{}
 \end{subfigure} 
\end{center}
\caption{(a) Two points are visible if their line-of-sight stays above the
  terrain. (b,c,d) Interpolating a grid terrain along the projection
  of the line-of-sight $(v,q)$.}
\label{fig:grid_LOS}
\end{figure*}

\subsection{Related work}

Viewsheds on terrains can be modeled in a continuous or discrete way.  
The continuous viewshed, commonly studied on polyhedral terrains, 
entails computing the exact visibility polygon; on triangulated
terrains it is known that its complexity can be $\Theta(n^2)$ in the
worst case~\cite{cole:visibility}.  For a survey of results on
continuous visibility we refer to~\cite{floriani:vissurvey}.
For the rest of the paper we'll consider a grid terrain $T$ of $n$
points, for simplicity assumed to have size $\sqrt n$-by-$\sqrt n$.
Viewsheds on grid terrains are modeled in a discrete way: the viewshed
of a viewpoint $v$ is defined as the set of all \emph{grid points}
that are visible from $v$, and is represented as a grid of the same
size as $T$. For simplicity we assume that each point is labeled as
either visible or not visible; algorithms can be easily extended to
compute for e.g. distance to visibility instead of binary visibility.

The straightforward algorithm to compute if a grid point $q$ is
visible from $v$ is to walk along the line-of-sight $vq$, compute the
intersections between the horizontal projection of $vq$ and the
horizontal and vertical gridlines, and compute the elevation of the
terrain at these points with linear interpolation; point $q$ is
visible if all these elevations are below $vq$
(see Figure~\ref{fig:grid_LOS}). This model, which we denote \emph{the gridlines
model}, corresponds to viewing the terrain as consisting of grid cells
bounded by the gridlines, with the behaviour of the terrain inside the
cells unspecified~\cite{floriani:vissurvey}.
A variant is to model the terrain as a triangular mesh by adding one
of the diagonals in each grid cell in the straightforward way;
computing if a grid point $q$ is visible from $v$ in this case comes
down to walking along $vq$ and finding the intersection of $vq$ with
the triangular faces, and computing the elevation of each intersection
point by interpolating linearly from the triangular face that contains
it. In fact, Larsen~\cite{larsen:viewshed} showed that the the
gridlines model is conceptually equivalent to adding in each cell the
diagonal that corresponds to the bottom of the tetrahedron spanned by
the four grid points, and thus the gridlines model is equivalent to a
triangular mesh.
In both variants, the line-of-sight $vq$ intersects $O(\sqrt n)$ grid
lines or triangles, respectively, and the visibility of a point $q$
can be determined in $O(\sqrt n)$ time. The straightforward algorithm
for computing the viewshed of $v$ is to iterate over all grid points
$q$ and check if they are visible from $v$; this runs in $\Theta(n
\sqrt n)$ time.
The gridlines model and this viewshed algorithm, referred to as \rrr,
were introduced by Franklin and Ray~\cite{franklin:sdh94}. \rrr is
considered to produce the ground truth viewshed
(e.g.~\cite{izraelevitz:vis}), so we use it as reference. For the rest
of this paper we denote the viewshed computed by \rrr as
viewshed$_T(v)$.

An improved approach for computing viewshed$_T(v)$ was described by
\\Haverkort et al~\cite{ht:visLI} and is based on the work of Cole and
Sharir~\cite{cole:visibility} and De Floriani and Magillo~\cite{floriani:horizon}. The idea
is to traverse the gridline segments in front-to-back order while
maintaining horizons. Put simply, the horizon $H_v$ of a set of
segments with respect to viewpoint $v$ is the upper rim of the
segments as they appear to a viewer at $v$. On a triangulated terrain
the horizon has complexity $O(n \cdot \alpha (n))$, where $n$ is the
number of vertices in the TIN and $\alpha()$ is the inverse of the
Ackerman function~\cite{cole:visibility}, while on a grid terrain it has
been shown to have  complexity $O(n)$~\cite{ht:visLI}. The
algorithms by Haverkort et al, which were described and analysed in
the IO-model but easily extend to the RAM model, traverse the grid in
layers around the viewpoint and compute horizons either in an
incremental or in a divide-and-conquer fashion. The two algorithms,
\visiter and \visdac, run in $O(n \sqrt n)$ and $O(n \lg n)$,
respectively.


The algorithms mentioned above are the only ones known for computing
viewshed$_T(v)$. Many other algorithms have been proposed that are
faster but compute an approximation of viewshed$_T(v)$. Van Kreveld's
radial sweep algorithm~\cite{kreveld:viewshed} runs in $O(n \lg n)$
time but assumes the terrain is as a tessellation of flat cells, where
each cell is centered around a grid point and has the same zenith
across its span. The use of the nearest-neighbor interpolation rather
than linear means that the algorithm computes a different viewshed
than \rrr, and unfortunately the algorithm does not immediately extend
to using linear interpolation. The model is not ideal in practice as
it is not continuous, and some artifacts of the viewshed computed by
this algorithm were reported in~\cite{ht:visLI}.
%
%
Several other viewshed algorithms run in $O(n)$ time by using
approximation. For e.g. \textsc{R2} by Franklin and
Ray~\cite{franklin:sdh94} considers the line-of-sights to only the
points on the boundary of the grid;
\textsc{XDraw}~\cite{franklin:sdh94} and its generalization
\textsc{BackTrack} by \cite{izraelevitz:vis} use an approximate
horizon given by the previous layers in order to determine if a point
$p$ is visible. For a survey of approximate algorithms we refer
to~\cite{floriani:vissurvey,ht:terrainsurvey2014}.

In addition to viewsheds, another visibility concept defined on grid
terrains is the \emph{total viewshed}: the total viewshed of a grid
terrain $T$ is another grid of the same size as $T$ whose value
at a grid point $v$ is the size of viewshed$_T(v)$, i.e. the number of
grid points in $T$ visible from $v$. 
The straightforward algorithm for computing the total viewshed is to
compute viewshed$_T(v)$ for every grid point~$v$; this runs in $O(n^2
\sqrt n)$ and $O(n^2 \lg n)$ time respectively using one of the
viewshed algorithms mentioned above. No total viewshed algorithm is
known that improves this quadratic bound, and a quadratic bound means
that the algorithm is unlikely to be feasible in practice on large
grids.
Recent progress has been achieved for the 1D-version of this problem
by~Afshani et al~\cite{lambrechts:totalview}. Their algorithm relies
on a clever mapping of 1D visibility  to the problem of
red-blue intersection counting in the dual plane, and achieves a bound
of $O(n \lg ^2n)$. Unfortunately the approach does not extend to 2D.

Recent research has focused on approximate algorithms and
parallelization in order to exploit the availability of multi-core
computers, clusters and GPUs (graphics processing units).  Parallel
implementations of \rrr, \textsc{R2}, and Van Kreveld's algorithms
were described by Zhao et al~\cite{zhao2013:parallelr3}, Osterman et
al~\cite{osterman2014:parallelr2} and Ferreira et
al~\cite{ferreira2014:parallelkreveld}, respectively.  Zhao et
al~\cite{zhao2013:parallelr3} target terrains that do not fit in main
memory, and report a factor of 50 speedup using GPUs. Ferreira et
al~\cite{ferreira2014:parallelkreveld} report speedup up to 12 on a
platform with 16 processors assuming datasets that fit in memory.
Osterman et al~\cite{osterman2014:parallelr2} report an implementation
of R2 that's faster than TiledVS~\cite{ferreira:vis}, and a factor of
10 speedup using GPUs, running on datasets up to 180GB.
Faster total viewshed algorithms can be obtained by using one of the
approximate viewshed algorithms. Tabik et al~\cite{tabik:2014}
described a different approximate total viewshed algorithm and a
parallel implementation based on discretizing the radial angle around
the viewpoint
and re-using the information in sectors between adjacent points.


\subsection{Our contributions} 

We describe a novel algorithm for computing viewshed$_T(v)$ using a
multiresolution approach. 
%
Given a parameter $k >1$ which represents the block size, our
algorithm starts by creating a grid~$T'$, which is a lower-resolution
version of ~$T$, such that each point in $T'$ corresponds to a block of
$\lceil \sqrt k \rceil \times \lceil \sqrt k \rceil = O(k)$ points in
$T$. This step
runs in $O(n)$ time and the resulting grid $T'$ has size
$\Theta(n/k)$, where $n$ denotes the size of the original grid, $T$.

Once $T'$ is created, our algorithm computes viewshed$_T(v)$ in two
steps:
The first step, \lores, which we describe in Section~\ref{sec:lores},
computes the viewshed of $v$ on $T'$ while maintaining the invariant
that any point in $T'$ that is labeled as invisible corresponds to a
block in $T$ that is guaranteed to not contain any visible points.
Thus, the role of this first step is to use the low-resolution
representation~$T'$ to filter out blocks that are guaranteed to be
invisible in $T$. The first step runs in $O(\frac nk \lg \frac nk)$.

The second step, \hires, which we describe in Section~\ref{sec:hires},
starts with  the blocks that were marked visible by \lores in $T'$ and
computes the visibility of their points using the full resolution data
in $T$. The complexity of this step is $O(l \cdot \lg n + \frac nk + k
\lg k + n )$, where $l$
is the
total size of blocks marked visible by \lores.

Overall our algorithm runs in $O(l \cdot \lg n + \frac nk \lg \frac nk
+ k \lg k + n )$. In practice, for values of the block size $k$ such
that $ k = \Omega(1)$, and if $l << n $, the running time improves on
the previous best bound of $O(n \lg n)$~\cite{ht:visLI}.
Using this algorithm we can compute the total viewshed in $O(n \cdot
(\frac nk \lg \frac nk + k \lg k + l\cdot \lg n))$; when $k$ is large and $l <<
n$ this bound may be sub-quadratic and improves the previous bound of
$O(n^2 \lg n)$.

In Section~\ref{sec:exper} we describe experimental results showing
the behavior of our algorithm in practice and a comparison with the
previous algorithms--- the algorithm from~\cite{ht:visLI} and an
implementation of \rrr~\cite{franklin:sdh94}. We discuss our findings
regarding the choice of block size and its impact on performance. Our
new algorithm achieves a speedup of 20 or more on large,
high-resolution grids. 
%
In Section~\ref{sec:concl} we comment on our results and discuss ideas
for future work.

\begin{figure*}[t]
  \begin{center} 
    \begin{subfigure}{.3\textwidth} 
      \centering{
        \includegraphics[width=.9\linewidth]{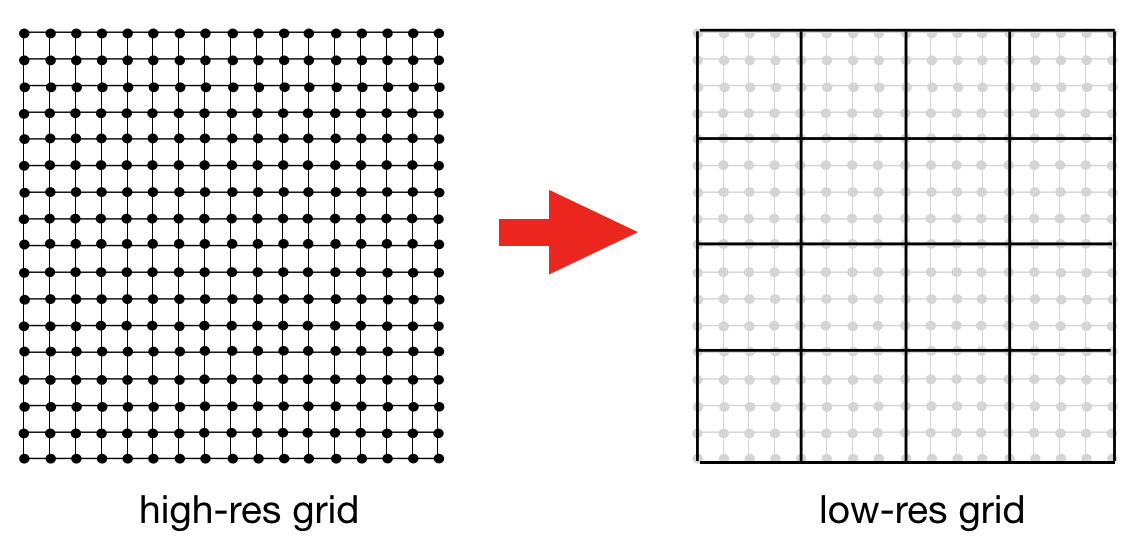}
        \caption{}
      }
    \end{subfigure} 
    \begin{subfigure}{.3\textwidth} 
      \centering{
        \includegraphics[width=.9\linewidth]{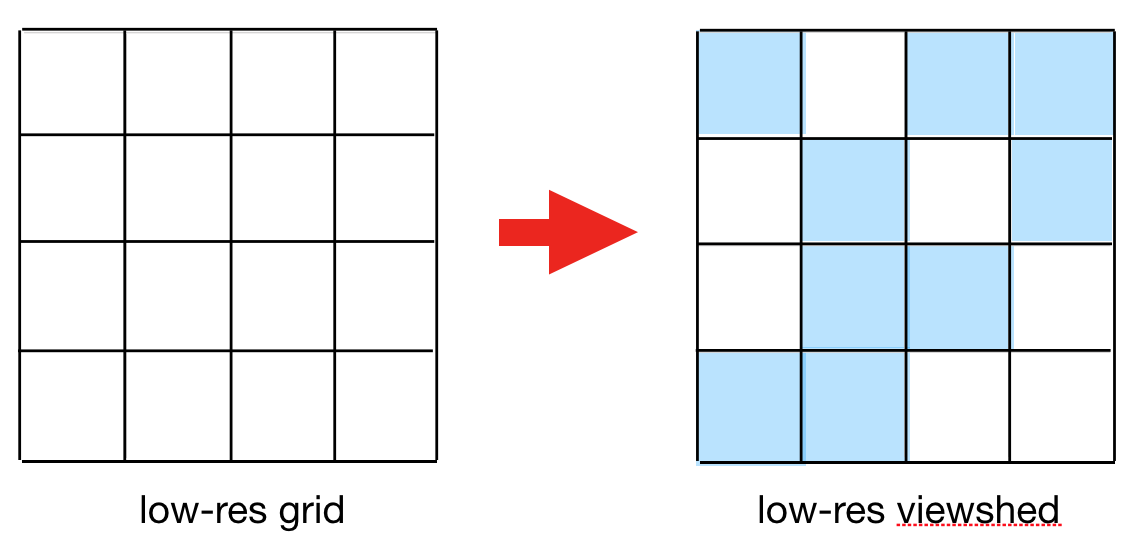}
        \caption{}
      }
    \end{subfigure} 
    \begin{subfigure}{.3\textwidth} 
      \centering{
        \includegraphics[width=.9\linewidth]{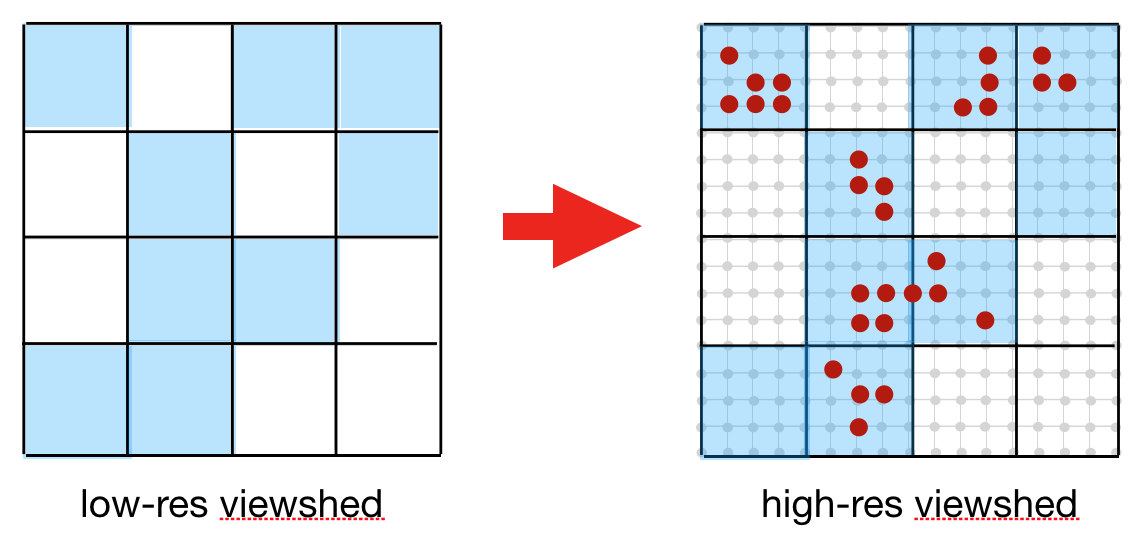}
        \caption{}
      }
    \end{subfigure} 
  \end{center}

\caption{ The outline of our algorithm. (a) Create the low-res. grid  (b) Compute the low-res. viewshed (c) Compute the high-res. viewshed}
\label{fig:outline}
\end{figure*}

\section{Notation and Background}

Let $T$ be a terrain represented as an elevation grid. For ease of presentation we assume the grid is square and has size $\sqrt n$~by~$\sqrt n$. The grid is given as a matrix $Z$, stored row by row, where $Z_{ij}$ is the elevation of the point in row $i$ and column $j$. The viewshed of $v$ is stored as a matrix $V$ of the same size as $Z$, in which $V_{ij}$ is 1 if the point in row $i$ and column $j$ is visible from $v$, and 0 otherwise.

We consider a spherical coordinate system centered at the viewpoint
$v$. 
The azimuth angle of a point $p=(p_x, p_y,p_z)$ with respect to
$v=(v_x, v_y, v_z)$ is the angle of the projection of $p$ onto the x-y
plane, measured counterclockwise from the positive x-axis when viewed
from above.
The zenith angle of a point $p=(p_x, p_y,p_z)$ with respect to
$v=(v_x, v_y, v_z)$ is the complement of the angle of $vp$ with the
positive z-axis: zenith$(p) = \texttt{arctan} {(p_z -
  v_z)}/ {dist(p,v)}$, where $dist(p,v)$ is the Euclidean distance
between $p$ and $v$ in the xy-plane, $dist(p,v) = | (p_x, p_y) - (v_x,
v_y)| = ((p_x -v_x)^2 + (p_y -  v_y))^{1/2}$. 
A point $p$ is visible from $v$ if and only if the zenith of $p$ is
higher than the zenith of any point on $T$ whose projection lies on
the line segment from $(v_x,v_y)$ to $(p_x,p_y)$.

We view the terrain as a consisting of 3D square cells bounded by
gridlines, that is, $T$ consists of the set of segments of the
form $((i,j,Z_{ij}), (i, j+1,Z_{i,j+1}))$  for  all $ 0\le i <\sqrt n,
0 \le j < \sqrt n -1$ and $((i,j,Z_{ij}), (i+1,
j,Z_{i+1,j}))$,  for  all $0 \le i < \sqrt n-1, 0 \le j < \sqrt n$.
%

Our algorithms use the concepts of front-to-back ordering and horizon
based on Cole and Sharir~\cite{cole:visibility}. Given a set $S$ of
segments in the plane, we say that a segment $(a,b)$ is \emph{in
  front of} a segment $(c,d)$, $(a,b)\prec(c,d)$ with respect to $v$,
if there exists a horizontal ray starting at $v$ that intersects both
$(a,b)$ and $(c,d)$ and the intersection of the ray with $(a,b)$ is
closer to $v$ than the intersection of the ray with $(c,d)$. When the
set of segments correspond to a grid, the in-front relation $\prec$
constitutes a partial order. A \emph{front-to-back order} of $S$ is
defined as a (total) ordering of the segments in $S$ that extends the
in-front partial order.

The horizon of a set $T$ of 3D segments with respect to a viewpoint
$v$ is a function $h(\theta)$, defined for any radial direction
$\theta \in [0, 2 \pi)$ as follows: $h(\theta)$ is the maximum zenith
of all points $p$ in $T$ whose azimuth is $\theta$. 
Intuitively, the horizon gives, for each radial direction $\theta$,
the boundary between the terrain and the sky.
%
The horizon of $v$ is equivalent to the upper envelope of the segments
in $T$ expressed in spherical coordinates and projected onto an
infinite vertical cylinder centered at $v$.

Many algorithms that compute visibility on triangulated terrains
traverse the edges of the terrain in front-to-back order of their
projections on the xy-plane with respect to the viewpoint $v$ and
compute horizons~\cite{floriani:vissurvey}.
Consider a segment $(a,b)$ in $T$ and let $H$ be the horizon of the
segments that come before $(a,b)$ in front-to-back order. When
traversing the segment $(a,b)$, all segments of $T$ whose projections
on the horizontal plane intersect the projection of $(a,b)$ come
before $(a,b)$ in front-to-back order, and therefore are in $H$. When
traversing segment $(a,b)$, the visibility of points $a$ and $b$ can
be determined by checking $a$ and $b$ against $H$ to see if they are
above or below, and based on that are labeled as visible or invisible,
respectively. The segments in $T$ are traversed in fron-to-back order,
either incrementally or using
divide-and-conquer~\cite{cole:visibility}. With the divide-and-conquer
approach the idea is to split the segments into two halves, compute
visibility and horizons recursively in each half, filter the points in
the second half by the horizon of the first half and merge the two
horizons.

The algorithm of Haverkort et al~\cite{ht:visLI} for computing
viewshed$_T(v)$ uses the same generic approach which is adapted to
exploit the regular structure of a grid. The algorithms in this paper
are based on it so we briefly review it. For $l \ge 0$, let
\emph{layer} $l$ of the grid, denoted $L_l$, denote the set of grid
points whose $d_{\infty}$-distance from the viewpoint, measured in the
horizontal plane, is $l$. By definition, $L_0$ consists of one point,
$v$. The algorithm in~\cite{ht:visLI} traverses the grid in layers
around the viewpoint, starting from $L_1$ and going outward; points in
the same layer are traversed in order of their azimuth around $v$.
This constitutes a front-to-back order as defined above. The
visibility of a point in layer $i$ is determined based on the horizon
$H_{\le i-1}$ formed by the previous $i-1$ layers. After the
visibility of all points in layer $i$ has been determined, the horizon
$H_i$ of points in layer $i$ is computed and merged with the horizon
of the previous layers, $H_{\le i-1}$, to form the new horizon $H_{\le
  i}$ of points in layers $1$ through $i$. This algorithm can be
implemented in an incremental or divide-and-conquer fashion.
On a grid of $n$ points, layer~$i$ has $O(i)$ points and the number of
points in layers $1$ through $i$ is $O(i^2)$. The horizon of layer $i$
has size $O(i) = O(\sqrt n)$ and $H_{\le i}$ has size $O(i^2) = O(n)$.
Traversing one layer, computing its visibility, computing its horizon
and merging it with the previous horizon takes $O(\sqrt n + n) = O(n)$
time. Since there are $O(\sqrt n)$ layers an incremental algorithm
runs in $O(n \sqrt n)$. This can be improved using divide-and-conquer
to $O(n \lg n)$.

\section{Description of Our Algorithm}
\label{sec:algorithm}

This section describes our new algorithm for computing
viewshed$_T(v)$. Its idea is rooted in the observation that viewsheds
in practice are small: on all terrains that we tried, including
regions that are relatively flat, the average viewshed size was below
$5\%$ of the size of the terrain\footnote{Although the average size
  of a viewshed is small, a viewshed can be
  spread out over a large area around $v$ so heuristic approaches that
  filter out points that are far away from $v$ will miss part of the
  visible region.}.
Since viewsheds are small, high resolution data is necessary to
compute precisely which points are visible. On the other hand, since
95\% of the terrain is not visible, low-resolution data may be
sufficient to identify regions that are not visible.
Our idea is to create a grid $T'$ which is a lower-resolution version
of $T$ and use it as a shortcut to 
identify blocks that are guaranteed to be invisible.

\subsection{ Outline} 
\label{sec:blockgrid} 

{\bf The low-resolution grid:} Given a terrain $T$ represented by a
grid of size $\sqrt n$ by $\sqrt n$ and a parameter $k >1$ which
represents the block size, the first step in our algorithm is to
create a grid $T'$ by partitioning $T$ into blocks of size $\sqrt
k$-by-$\sqrt k$; Each point in $T'$ corresponds to a block in $T$ and
stores two values: the maximum and minimum elevation in that block.
Grid $T'$ has size $O(n/k)$ and can be created in $O(n)$ time in a
straightforward way: traverse the original grid $T$, and for each
point $p=(i,j)$, compute the block $(i',j')$ that contains $p$, and
update the maximum and minimum elevation of block $(i',j')$ in $T'$
based on $Z_{ij}$.
For the rest of the paper we refer to $T'$ as the low-resolution grid
and to $T$ as the high-resolution grid (because it contains all the
data). Each ``point'' in $T'$ corresponds to a block in $T$, so we'll
refer to the points in $T'$ as blocks.

\vspace{.2cm}
\noindent {\bf \lores and \hires: } Given $T$, $T'$ and a viewpoint
$v$, our algorithm for computing viewshed$_T(v)$ has two steps: The
first step, \lores, operates on $T'$ and computes the viewshed of $v$
on $T'$, which we refer as the low-resolution viewshed; this is a grid
of the same size as $T'$ in which each block is marked as being
visible or invisible from $v$. The second step, \hires, considers the
blocks marked as visible by \lores and computes the visibility of
their points with full accuracy using the data from $T$. The blocks
that are marked as invisible by \lores are not examined further by
\hires. We guarantee that there is no loss of precision by enforcing
that any block that is marked invisible by \lores is guaranteed to not
contain any visible points. Put differently, the low-resolution
viewshed computed by \lores on $T'$ must be a superset of the viewshed
on $T$.
We describe the two steps in detail below.

\subsection{\lores: The low-resolution viewshed}
\label{sec:lores}

\begin{figure}[t]
  \begin{center} 
 \begin{subfigure}{.45\textwidth} 
      \centering{
        \includegraphics[width=.6\linewidth]{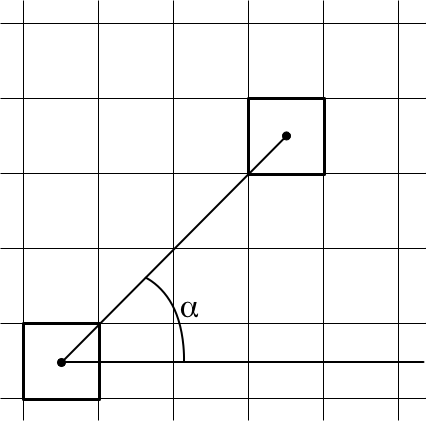}
        \caption{The azimuth of a block, $azimuth(B) = \alpha$}
      }
    \end{subfigure} 
    \begin{subfigure}{.45\textwidth} 
      \centering{
        \includegraphics[width=.6\linewidth]{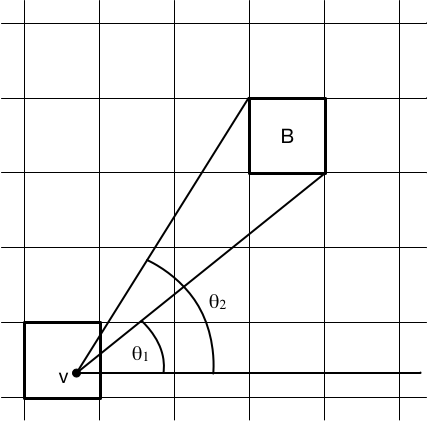}
        \caption{The span of a block $span(B) = [\Theta_1, \Theta_2]$}
      }
    \end{subfigure} 
  \end{center} 
  \caption{Block azimuth and block span}
  \label{fig:block_parameters}
\end{figure}

Given a grid terrain $T$, a low-resolution grid $T'$ as described
above, and a viewpoint $v=(r_v,c_v)$ in $T$, we describe how \lores
computes the low-resolution viewshed of $v$ on $T'$ while maintaining
the invariant that the blocks of $T'$ that it labels as invisible are
guaranteed to not contain any visible points.

We start with our notation. Denote by $B_v$ the block in $T'$ that
contains~$v$,
 and let $B$ be an arbitrary block, $B \neq B_v$. We define the
 following: 

{\bf Block azimuth:} The azimuth of a block $B$, $azimuth(B)$,
  is the counter-clockwise
  angle in the horizontal plane between the line connecting the center
  of $B_v$ to the center
  of $B$,  and the ray from the center of $B_v$ parallel to the
  $x$-axis; see Fig.~\ref{fig:block_parameters}(a). 

  {\bf Block span:} The span of a block $B$,
  $span(B)=[\theta_1,\theta_2]$, is the
  interval from the minimum azimuth $\theta_1$ of a point
  contained in
  $B$ to the maximum azimuth $\theta_2$ of a point contained in
  $B$; see Fig.~\ref{fig:block_parameters}(b). 

  {\bf Block minimum zenith:} The minimum zenith $z_{min}(B)$ of a
  block $B$ is the minimum
  possible zenith of a point contained in $B$. More precisely, let
  $m=(r_m,c_m,Z_m)$ be the grid point of minimum elevation in $B$ and
  $d_{min}$ and $d_{max}$ be the minimum and maximum distances from
  $v$ to a point in $B$, respectively. If $Z_m>Z_v$, then
  $z_{min}(B)=\arctan((Z_m-Z_v)/d_{max})$. Otherwise,
  $z_{min}(B)=\arctan((Z_v - Z_m)/d_{min})$.

  {\bf Block maximum zenith:} Similarly, we define the maximum
  zenith $z_{max}(B)$ of a block $B$ as the maximum possible zenith of
  a point contained in $B$.
 Let $m=(r_m,c_m, Z_m)$ be the grid point of maximum elevation in $B$ and
 $d_{min}$ and $d_{max}$ be the minimum and maximum distances from $v$
 to a point in $B$, respectively. If $Z_m>Z_v$, then
 $z_{max}(B)=\arctan((Z_m-Z_v)/d_{min})$. Otherwise,
 $z_{max}(B)=\arctan((Z_v - Z_m)/d_{max})$.

 {\bf Block in-front relation:} Recall the in-front relation
 defined on a set of segments in the plane: a segment $(a,b)$ is in
 front of a segment $(c,d)$ with respect to $v$ , if there exists a
 horizontal ray originating at $v$ that intersects both $(a,b)$ and
 $(c,d)$ and the intersection of the ray with $(a,b)$ is closer to $v$
 than the intersection with $(c,d)$.
 We define an in-front relation on blocks as follows:
 Given two blocks $A,B$ in the plane, we say that $A$ is in front of
 $B$, $A\prec B$, if there exists a horizontal ray starting at $v$
 that intersects both $A$ and $B$ and the intersection with $A$ is
 closer to $v$ than the intersection with $B$.

\begin{figure*}[t]
  \begin{center} 
 \begin{subfigure}{.3\textwidth} 
      \centering{
        \includegraphics[width=.9\linewidth]{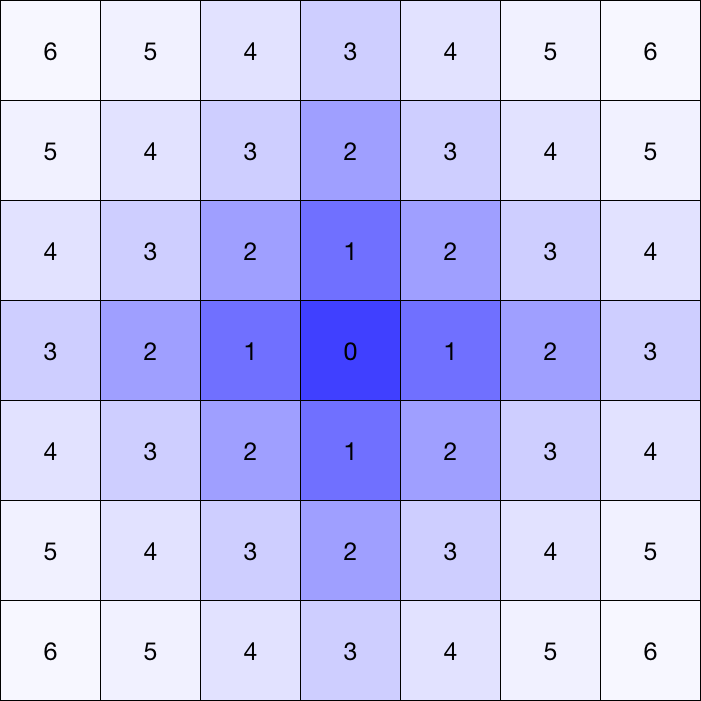}
        \caption{Block Manhattan distance}
      }
    \end{subfigure} 
    \begin{subfigure}{.3\textwidth} 
      \centering{
        \includegraphics[width=.9\linewidth]{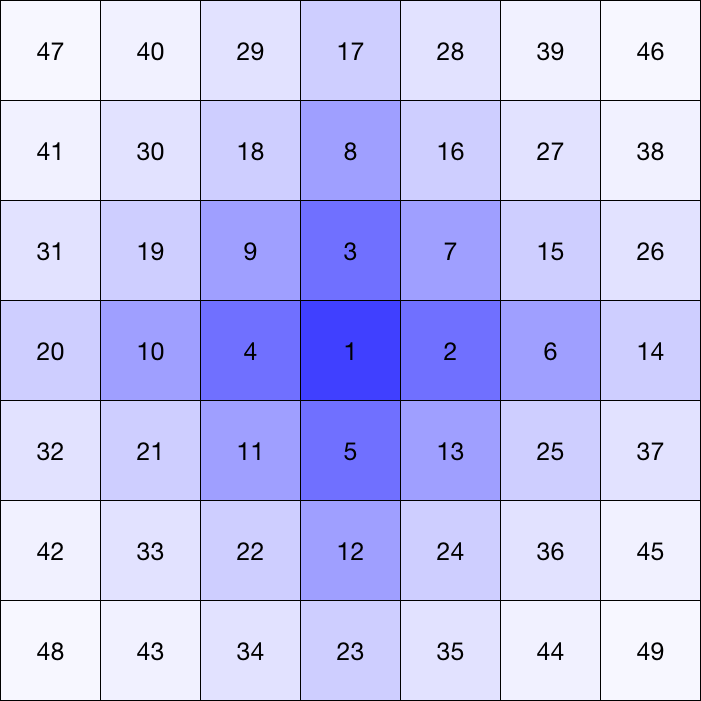}
        \caption{\lores block total order}
      }
    \end{subfigure} 
\begin{subfigure}{.3\textwidth} 
      \centering{
        \includegraphics[width=.9\linewidth]{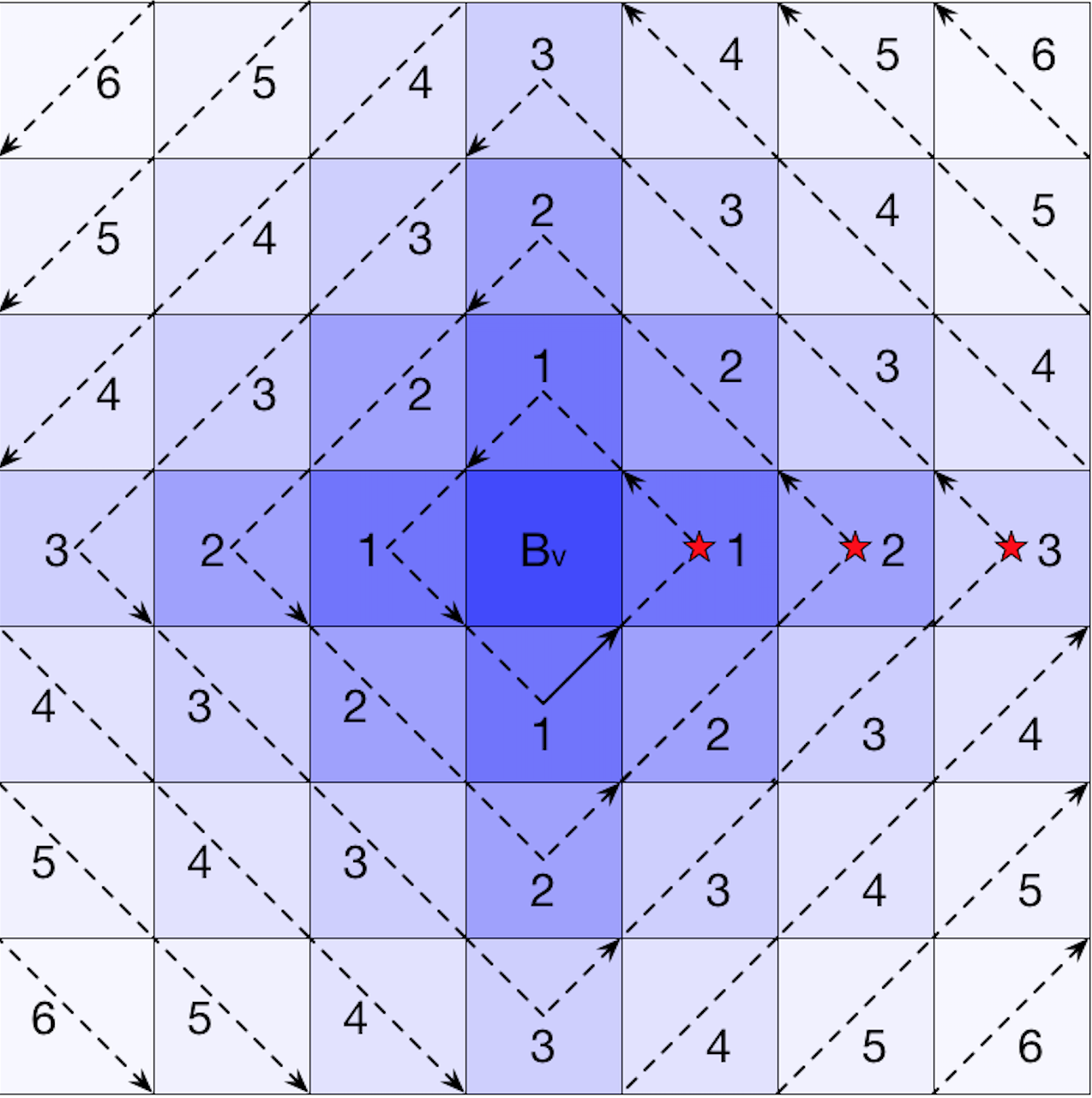}
        \caption{\lores block layers}
      }
    \end{subfigure} 
  \end{center} 
  \caption{\lores block traversal  order: blocks are traversed in order of
    increasing Manhattan distance from $B_v$, and inside a layer, in
    order of block azimuth.}
  \label{fig:block_order}
\end{figure*}

 {\bf Block front-to-back order:} Given two blocks $A$ and $B$ at
 coordinates
 $(r_A, c_A)$ and $(r_B, c_B)$ respectively in $T'$, we denote
 $dist_1(A,B) = |r_A - r_B| + |c_A - c_B|$ the Manhattan distance
 between $A$ and $B$. Figure~\ref{fig:block_order}(a) shows the
 Manhattan distances of blocks with respect to block $B_v$ which is in
 the center.
 We use the following total order, which can be
 easily verified that it extends the in-front partial order defined
 above:
%
 We say that $A<B$ if $dist_1(A,B_v)<dist_1(B,B_v)$ or if $dist_1(A,B_v)=dist_1(B,B_v)$
 and $azimuth(A) < azimuth(B)$. 
 It is illustrated in Figure\ref{fig:block_order} (b), with
 block $B_v$ starting at index $1$.

 {\bf Block layers:} A block $B$ is said to be in layer $l$ if 
 $dist_1(B, B_v) = l$. For $l \ge 0$, block layer $l$ contains all
 blocks $B$ in $T'$ such that $dist_1(B,B_v)=l$; see Figure
 \ref{fig:block_order}(c). Traversing the blocks in $T'$ in
 front-to-back order as defined above corresponds to traversing the
 grid $T'$ in layers.

\vspace{.5\baselineskip} \lores is based on a front-to-back traversal
of $T'$ while maintaining block layer horizons. The difference to previous
algorithms
is that the elements in $T'$ represent blocks rather than points. Each
block $B$ in $T'$ stores its minimum and maximum elevation, therefore
$z_{min}(B)$, $z_{max}(B)$, $span(B)$ and $azimuth(B)$ can all be
computed on the fly in $O(1)$ time.

The invariant that \lores must maintain is that any block that it
marks as invisible is guaranteed to not contain any points in
viewshed$_T(v)$. \lores maintains this invariant by using the minimum
and maximum zenith of a block to get lower and upper bounds for the
horizon of the points in that block. Let $H$ be the horizon formed by
the first $i-1$ layers of blocks, that is, the horizon of all points
in $T$ contained in blocks $B'$ s.th. $dist_1(B',B_v)\leq i-1$;
consider a block $B$ on layer $i$, that is, $dist_1(B,B_v) = i$.
\lores relies on two observations:

\begin{lemma} 
\label{lemma1}
  If $z_{max}(B)<H(\theta)$ for all $\theta \in span(B)$, then
  all the points in $B$ are guaranteed to be invisible in $T$.
\end{lemma}
Proof: For any point $p$ in $B$, its zenith
  $z_p$ is upper-bounded by $z_{max}(B)$, and therefore $z_p <
  z_{max}(B) < H$, thus $p$ is invisible.

   Lemma \ref{lemma1} gives a sufficient condition for the points in a block $B$
   to be invisible, provided we know the horizon $H$ of
   the blocks in front of $B$. Since 
   $T'$ does not have knowledge of the exact horizon of a block, 
   we use the minimum zenith of a block as a lower bound for the
   horizon of a block, which gives a lower bound for $H$. We have:

\begin{lemma}
  \label{lemma2} Let $H'$
  be the horizon obtained by inserting each block $B'$ such that $B' <
  B$ as a segment with the same span as $B'$ and zenith equal to
  $z_{min}(B')$. Then $H \ge H'$.
\end{lemma}

Proof: For any point $p$ in $B'$, its zenith $z_p$ is lower-bounded by
$z_{min}(B)$, and therefore the horizon of $B'$ is lower-bounded by
$z_{min}(B)$.
Therefore $ H = \bigcup_{} H(B') \ge \bigcup  z_{min}(B') = H'$. 

%
%

Putting Lemma \ref{lemma1} and \ref{lemma2} together, if a segment
with the same span as $B$ and zenith equal to $z_{max}(B)$ lies below
$H'$, then it lies below $H$ and thus all points in $B$ are guaranteed
to be invisible.

The overall \lores algorithm is given in Figure~\ref{fig:lores_alg}.
It is an incremental algorithm, where the blocks are traversed in
front-to-back order as defined above: layer by layer, starting from
layer $i=1$ and, inside a layer, in order of increasing block azimuth.
The visibility of a block $B$ in layer $i$ is determined based on the
(lower bound) horizon $H$ of blocks in layers $1$ through~$i-1$: by
Lemma\ref{lemma1} we know that block $B$ is invisible if $z_{max}(B)$
is below $H$ for all $\theta \in span(B)$. Thus, checking whether a
block is invisible reduces to checking whether a segment is completely
below the horizon $H$; in Figure~\ref{fig:lores_alg} this is
encapsulated in the function $\proc{IsInvisible} (B,H)$. A block is
considered visible if it's not invisible.
After the visibility of $B$ is determined, it is inserted into the
horizon as a segment with the same  span as  $B$,  at zenith equal to
$z_{min}(B)$; in Figure~\ref{fig:lores_alg} this is
encapsulated in the function $\proc{HorizonInsert} (B,H_{tmp})$. 
For efficiency reasons, rather than inserting one block at a time in
$H$, we build the horizon of all blocks in layer $i$ in a layer
horizon which is denoted by $H_{tmp}$ in Figure~\ref{fig:lores_alg}. Once
all the blocks in layer $i$ have been traversed, $H$ and $H_{tmp}$ are
merged to form the new horizon.

\begin{figure*}[t]
  \centering
  \fbox{
    \parbox{12cm}{
      \begin{codebox}
        \Procname{ $\proc{LoRes} (T', r_v, c_v)$}
        \li \Comment Input: low-res. grid $T'$, viewpoint coord.
        $(r_v, c_v)$
        \li \Comment Output:  low-res. viewshed grid $V_{T'}$
        
        \li $H=\emptyset$
        \li Create a grid $V_{T'}$ of same size as $T'$
        \li \For $i=1$; $i\leq$ number of block layers in $T'$; $i++$ \Do
        \li $H_{tmp}=\emptyset$
        \li \For block $B$ in layer $i$ in order of increasing block azimuth \Do
        \li set  $B$ as  $\proc{IsInvisible} (B,H)$? \invisible: \visible
        \li $\proc{HorizonInsert} (B,H_{tmp})$
        \End
        \li H = $\proc{MergeHorizons} (H,H_{tmp})$
        \End
\li return grid $V_{T'}$
      \end{codebox}
    }
  }
  \caption{\lores algorithm (incremental)}
  \label{fig:lores_alg}
\end{figure*}

\paragraph{\lores Analysis} Denote the number of blocks in $T'$ by
$n'=\Theta(\frac{n}{k})$. In order to determine if a block $B$ is visible,
the portion of $H$ containing $span(B)$ must be traversed. Layer $i$
is traversed in order of increasing azimuth, and $H$ is stored in
order of increasing azimuth so in order to determine the visibility of
all the blocks in layer $i$, the horizon $H$ must be traversed exactly
once (this is the reason that our algorithm is incremental at the
layer level, not block level). Similarly, each block is inserted at
the end of $H_{tmp}$ in $O(1)$ time.
Layer $i$  has $\Theta(i)$ blocks and the horizon of layer $i$,
$H_{tmp}$, has size $O(i)$. Finally, in order to merge $H_{tmp}$ and
$H$, both horizons must be traversed exactly once, which takes
$O(i+|H|)$ time. Therefore, the running time necessary to determine
the visibility of blocks in layer $i$ and to merge $H_{tmp}$ and $H$
is $O(i+|H|)$.
%
Haverkort et al~\cite{ht:visLI} showed that the horizon of a grid of
$n'$ points has complexity $O(n')$; based on this we have that $|H| = O(n')$.
Summing all the layers, the total running time of the incremental
approach 
 is $\sum_{i=1}^{O(\sqrt{n'})}(O(i) + |H|) = O(n' \sqrt n')$.

\paragraph{Divide-and-conquer \lores} 
The running time of the incremental algorithm in
Figure~\ref{fig:lores_alg} can be improved to $O(n'\log n')$ using
divide-and-conquer as in~\cite{cole:visibility}.
Consider the $n'$ blocks in front-to-back order as defined above.
Instead of traversing them one layer at a time, they are divided in
two halves, where the first half contains all the blocks that come
before the blocks in the second half. We compute visibility and
horizons for the first and second half of the blocks recursively. Let
$H_1$ and $H_2$ be the horizons for the first and second halves of the
blocks, respectively, which are returned by the recursive calls. In
the merge phase, we filter the second half blocks through $H_1$, and
merge $H_1$ and $H_2$ into a horizon $H$ and return it.
%
It takes $|H_1|+O(n')+|H_2|$ time to update the visibility of blocks
in $H_2$ using $H_1$ and to merge $H_1$ and $H_2$. Overall the running
time is given by $T(n')=2T(\frac{n'}{2})+|H_1|+O(n')+|H_2|$. Since
$|H_1|=|H_2|=O(\frac{n'}{2})$, this solve to $O(n'\log n')$. We have
the following:

\begin{lemma} 
\label{lemma:lores} Given a grid terrain $T$ and a low-resolution
  representation $T'$ of size $\Theta (n/k)$ as described above, the
  low-resolution viewshed $V_{T'}$ can be computed
  in $O(n' \lg n')$ time, where $n' = \Theta(n/k)$. \end{lemma}

\subsection{\hires}
\label{sec:hires}

\begin{figure}[t]
   \begin{center} 
           \centering{
         \includegraphics[width=.4\linewidth]{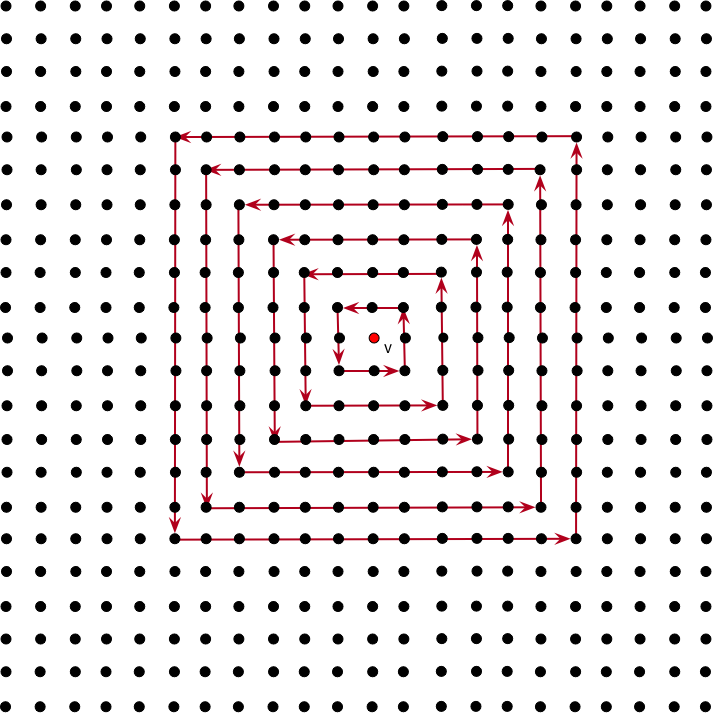}
         \caption{}
       }
       \caption{The algorithm traverses the  grid points in layers
       around $v$ defined based on the    $d_{\infty}$-distance.}
    \label{fig:visiter}
  \end{center} 
\end{figure}

    

This section describes the second step of our algorithm, \hires.
\hires takes as input the elevation grid~$T$, the row and column
coordinates $(r_v, c_v)$ of the viewpoint $v$ in $T$, and the
low-resolution viewshed $V_{T'}$ of $v$ created by \lores, which we
assume is a grid of the same size as $T'$ in which each block is
marked as visible or invisible from $v$. The output of \hires is
viewshed$_T(v)$, which is a grid of the same size as $T$, in which
each point is marked as visible or invisible from $v$; we denote it 
 $V_T$.

 

 \hires uses the same idea of a front-to-back traversal of the grid
 while computing horizons and is based on the algorithm
 in~\cite{ht:visLI}. The algorithm by Haverkort et. al~\cite{ht:visLI}
 traverses the grid in \emph{point layers} around $v$, which are
 defined based on the $d_{\infty}$-distance from $v$: For $i \ge 0$,
 point layer $i$ consists of all grid points in $T$ with
 $dist_{\infty}(p,v)=i$, where $dist_{\infty} (p,q) = max
 \{ |r_p-r_q|, |c_p-c_q| \} $; see~Figure~\ref{fig:visiter}. Inside a
 layer, the points are traversed in order of their azimuth with
 respect to $v$. When a point $p$ is traversed, the gridline segments
 $(p,q)$ adjacent to $p$, where $q$ comes before $p$ in this order,
 are inserted in the horizon. Overall, this ordering extends the
 in-front relation between grid line segments, which ensures that when
 a point $p$ is traversed and its visibility is determined, the
 horizon contains all gridline segments intersected by $vp$.

\begin{figure*}[t]
   \begin{center} 
     \begin{subfigure}{.3\textwidth} 
       \centering{
         \includegraphics[width=.9\linewidth]{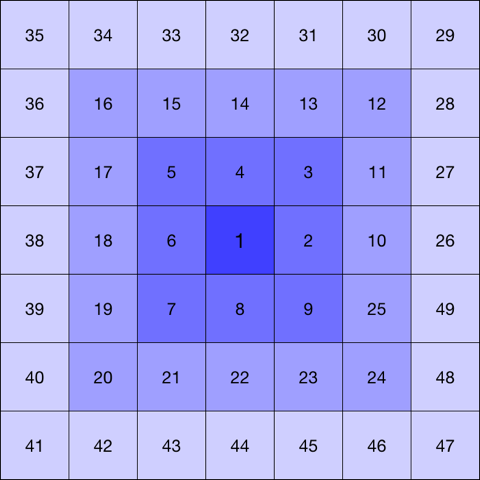}
         \caption{\hires block total order}
       }
     \end{subfigure} 
     \begin{subfigure}{.3\textwidth} 
       \centering{
         \includegraphics[width=.9\linewidth]{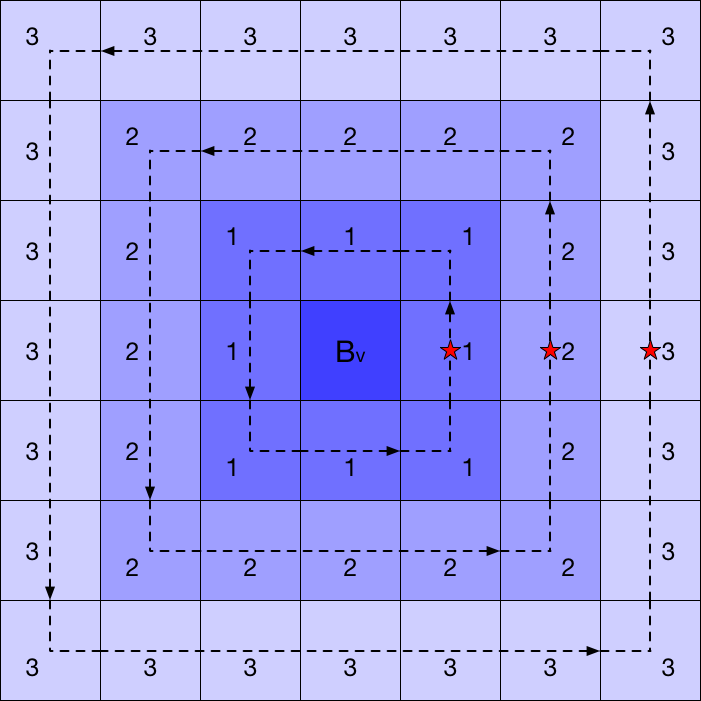}
         \caption{\hires block layers}
       }
     \end{subfigure} 
     \begin{subfigure}{.3\textwidth} 
       \centering{
\includegraphics[width=.9\linewidth]{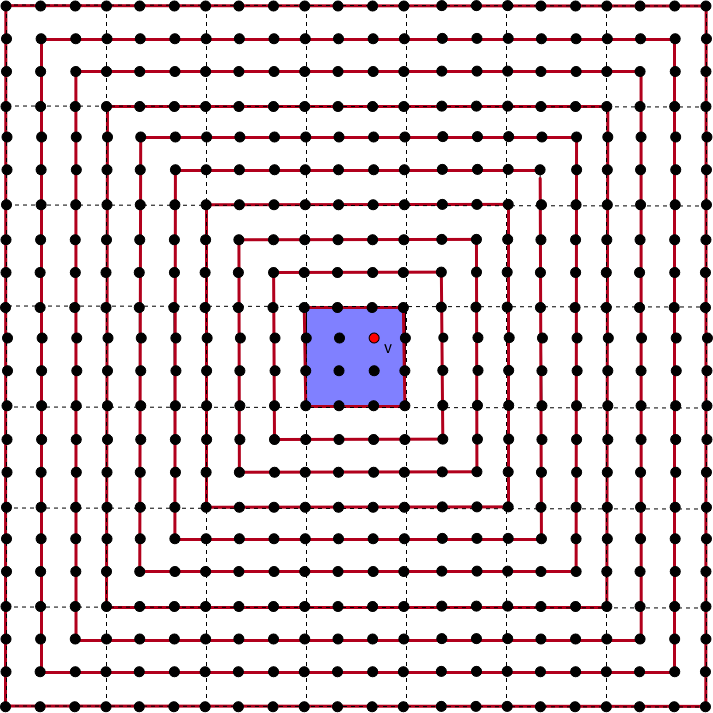}      
   \caption{\hires point layers}
       }
     \end{subfigure} 
     \caption{\hires traversal order: blocks and points are traversed
       in layers around $B_v$}
    \label{fig:hiresOrder}
  \end{center} 
\end{figure*}

\begin{figure*}[t]
   \begin{center} 
     \begin{subfigure}{.3\textwidth} 
       \centering{
         \includegraphics[width=.9\linewidth]{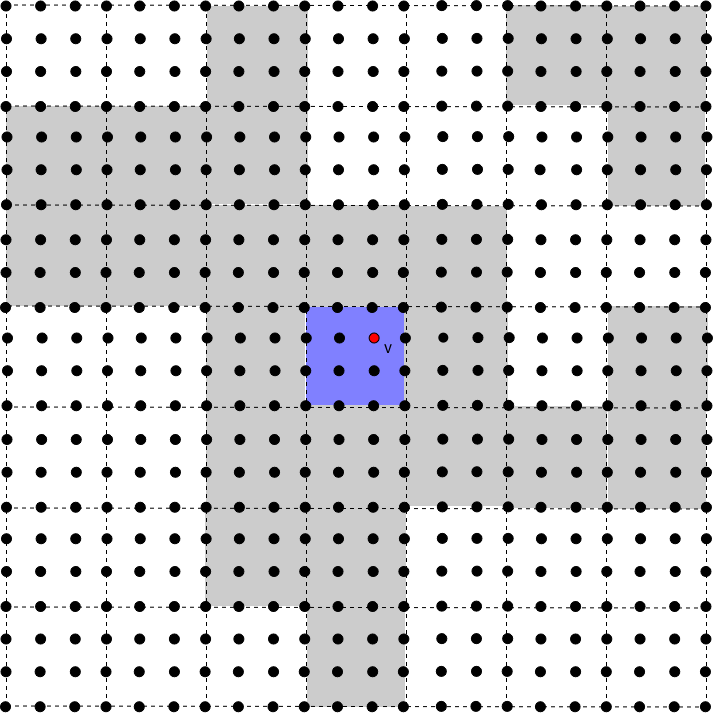}
         \caption{}
       }
     \end{subfigure} 
  \begin{subfigure}{.3\textwidth} 
       \centering{
         \includegraphics[width=.9\linewidth]{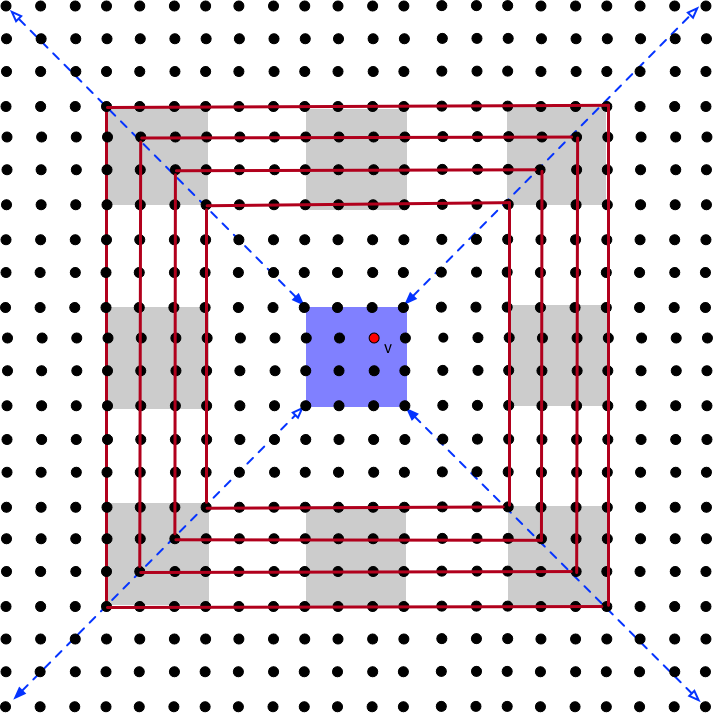}
   \caption{}
       }
     \end{subfigure} 
     \begin{subfigure}{.3\textwidth} 
       \centering{
         \includegraphics[width=.9\linewidth]{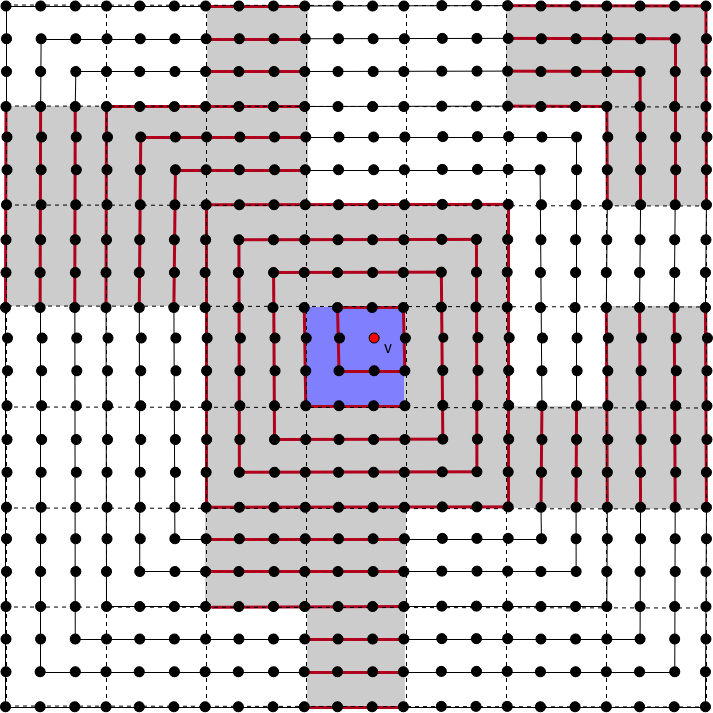}
         \caption{}
       }
     \end{subfigure} 

     \caption{\hires computes the viewshed on a union of blocks }
    \label{fig:hires}
  \end{center} 
\end{figure*}

Recall our notation: $k$ denotes the size of a block, $n'=\Theta(n/k)$
is the size of $T'$, $B_v$ is the block containing the viewpoint, and
$l$ is the total number of points in blocks that are visible in
$V_{T'}$. 
The goal for \hires is to compute visibility for a subset of points in
the grid corresponding to a union of blocks, rather than for the whole
grid. Figure \ref{fig:hires}(a) shows $B_v$, the visible blocks in
$V_{T'}$ (in gray), and the invisible blocks in $V_{T'}$ (in white).

Conceptually \hires has two steps: extract the points contained in
visible blocks in a front-to-back ordering (in our case: layer by
layer), and use the layers to compute their visibility.
To extract the point and build the layers, the straightforward idea
would be to traverse the original grid $T$ one layer at a time,
starting from the innermost and going outward; when reaching a point
$p$ that lies in a block that's marked as invisible in $V_{T'}$, skip
it; otherwise add $p$ to layer $i$, where $i=dist_{\infty} (p,v)$.
Since this requires reading the whole grid $T$, it runs in $O(n)$
which is not optimal when $l << n$.
This could be improved if, when reaching a point $p$ that falls in an
invisible block say $B$, we skip over the subsequent points in the
block and shortcut to the first point $p'$ that comes after $p$ which
is on the boundary of $B$; since each block has $O(\sqrt k)$ points on
its boundary, each invisible block would be encountered $O(\sqrt k)$
times (rather than $O(k)$ times above), which is an improvement, but
still not optimal.
In order to arrange the $l$ points into point layers in optimal $O(l)$
time, \hires starts from the low-resolution viewshed grid~$V_{T'}$: it
traverses the blocks in $V_{T'}$ and, when encountering a block marked
as visible, it read its $O(k)$ points from $T$ and inserts them into
their corresponding point layers. The main challenge is that the point
layers, as defined above, are centered at $v$, while blocks layers are
centered at $B_v$. 
To make block and point layer order consistent with each other we
define the following:

{\bf \hires block layers and block order:} For $i \ge 0$, block layer
$i$ in the low-resolution grid $V_{T'}$ contains all blocks $B$ with
$dist_\infty(B,B_v)=i$, where the $d_{\infty}$-distance between two
blocks $A(r_A, c_A)$ and $B(r_B, c_B)$ is $dist_{\infty} (A,B) = max
\{ |r_B-r_A|, |c_B-c_A| \}$. Inside a block layer the blocks are
ordered by increasing azimuth. The block layers used by \hires and the
corresponding block order are shown in Figure~\ref{fig:hiresOrder} (a,b).

{\bf \hires point layers:} \hires defines point layers with respect to
the boundary of block $B_v$ containing $v$:
For $i \ge 0$, point layer $i$ in $T$ consists of all grid points in
$T$ (and outside $B_v$) whose $d_{\infty}$-distance from the boundary of
$B_v$ is $i$. Point layer $0$ corresponds to the points on the
boundary of $B_v$; seeFigure~\ref{fig:hiresOrder} (c). Note that with
this definition a block layer maps directly to a set of point layers
and the other way around.

\begin{figure*}[t]
\centering
\fbox{
\parbox{\textwidth}{
\begin{codebox}
\Procname{$\proc{HiResVS}(T,V_{T'}, r_v, c_v)$}
\li \Comment Input: elev. grid $T$, low-res viewshed grid $V_{T'}$,
viewpoint $(r_v,c_v)$
\li \Comment Output: viewshed grid $V_T$
    \li $\proc{ComputeVisibility}(B_v)$
    \li $H = \proc{HorizonOf}(B_v)$

    \li Create $H_{tmp}=$ array of $\sqrt k$ horizons 
    \li \For  each block  layer $i$ in $V_{T'}$ starting at $i=1$ \Do 
    \li \For each block $B$ in layer $i$ in order of incr. azimuth \Do
    
    \li \If $B$ is visible in $V_{T'}$  \Do
    \li \Comment read all points of $B$ and insert in point layers   
    \li \For $j=0$; $j<\sqrt k$; $j++$ \Do
    
    \li traverse point layer $j$ in block $B$ and for each
    $p$ \Do 
    \li $\proc{HorizonInsert} (v,p,H_{tmp}[j])$
    \End     
    
    \End  
    \End  
    \End 
    
    \li \For $j=0$; $j<\sqrt k$; $j++$ \Do
    \li \For point $p$ in $H_{tmp}[j]$ \Do
    \li \If $\proc{IsVisible}(v,p,H)$: set $p=$ \visible 
    \Then 
    \li \Else set $p=$ \invisible  
    \End 
    \End  
    \li $H=\proc{MergeHorizons}(H,H_{tmp}[j])$
    \li $\proc{Clear}(H_{tmp}[j])$
    \End 
    
    \End 
    
    \li \Return $V_T$ 
    
 \end{codebox}
}}
\caption{\hires algorithm}
\label{fig:hires_algo}
\end{figure*}

The overall \hires algorithm is given in Figure~\ref{fig:hires_algo}.
It starts by computing the visibility of the points in block $B_v$ and
initializes the horizon $H$ as the horizon of $B_v$ (line 3:4 in
Figure~\ref{fig:hires_algo}). Inside block $B_v$ the point layers are
defined around $v$ as in~\cite{ht:visLI}. Throughout the algorithm $H$
represents the horizon of the points traversed so far. Then it creates
an array called $H_{tmp}$ which will be used to store the horizons
corresponding to the $\sqrt k$ point layers in a block layer (line 5).

Next, \hires traverses the blocks in $V_{T'}$ layer by layer, starting
from the innermost ($i=1$) and moving outward (line 6:7), using the
block layer order defined above and shown in Figure \ref{fig:hiresOrder}(a).
For each block $B$, \hires checks if it is visible in $V_{T'}$ (line
8). If $B$ is not visible, \hires advances to the next block in that
layer. If $B$ is visible, it reads the $O(k)$ points in $B$ from $T$
and inserts them into the horizons for the corresponding point layers
(lines 9:12).
%

\vspace{.5\baselineskip} 
\emph{Extracting the point layers in a block
  (lines 9:12 in Figure~\ref{fig:hires_algo}):} The points in a block
layer in $V_{T'}$ correspond to $\sqrt k$ point layers in $T$. When
traversing block layer $i$ in $V_{T'}$, the horizons of the~$\sqrt k$
point layers are kept in the array $H_{tmp}[]$, where $H_{tmp}[j]$
represents the horizon of point-layer~$j$. These horizons are local to
a block layer, and are reset before moving to the next block layer
(line~18).
Consider a block $B$. There are 3 cases: the block is vertical,
horizontal, or corner; refer to Figure~\ref{fig:hires}(b). If $B$ is
vertical and to the right of $B_v$, then the point layers correspond
to columns in $B$, entering $B$ through the south border, and exiting
through the north border; The other cases are similar.
For each visible block $B$, \hires determines the orientation of the
point layers based on the type of block (vertical, horizontal,
corner), and traverses the block starting with the inner-most point
layer $j=0$ and moving outward (line 12). The points in point layer
$j$ are traversed in order of azimuth and inserted in $H_{tmp}[j]$
(line 11:12); for each point $p$, this reduces to appending
$(\theta_p, z_p)$ to the end of $H_{tmp}[j]$, which is encapsulated in
function \textsc{HorizonInsert} (line 12).
%


\vspace{.5\baselineskip} {\it Determining visibility in a block (lines
  13:17 in Figure~\ref{fig:hires_algo}):} Once the blocks in layer $i$
in $V_{T'}$ are traversed and their points stored in $H_{tmp}[j]$, for
$j=0$ to $\sqrt k$, \hires uses them to determine visibility. It
traverses the horizons $H_{tmp}[j]$, starting with the horizon
corresponding to the innermost point layer $j=0$ and moving outward
(line 13). The points in $H_{tmp}[j]$ are traversed in the order in
which they were inserted, which is in order of increasing azimuth.
\hires maintains the cumulative horizon $H$ of all the point layers
traversed before $j$. Each point $p$ in point layer $j$ is checked
against $H$ to determine if it is visible (line 15), and is set to
\visible or \invisible in $V_T$. When the visibility of all the points
in point layer $j$ has been determined, the horizon $H_{tmp}[j]$ is
merged with $H$ (line 17), and $H_{tmp}[j]$ is cleared (line 18); this
means that the next layer of blocks (line 6) will start their point
layers from scratch.

\vspace{.5\baselineskip} {\it \hires analysis:} 
The time to initialize all points in $V_T$ to \invisible is $O(n)$.
Computing visibility and the horizon in $B_v$ can be done in $O(k \lg
k)$ using the algorithm in~\cite{ht:visLI}.
%
%
The point layers are created by traversing $V_{T'}$ and the points of
$T$ contained in visible blocks (lines 6:12); as we read the points
$p$ in a block, \textsc{HorizonInsert} can insert $p$ in the
corresponding point layer $j$ simply by appending it to the end of
$H_{tmp}[j]$; Overall this runs in linear time, or $O(n/k + l)$.
After a whole layer of blocks has been traversed and the layer
horizons $H_{tmp}[j]$, for $0 \le j < \sqrt n$, have been
populated, each $H_{tmp}[j]$ is traversed to determine visibility of
its points and then it is merged with the current horizon~$H$; Both
$H_{tmp}[j]$ and
  $H$ are stored in order of increasing azimuth, so determining
  visibility of all points in $H_{tmp}[j]$ and merging with $H$ can be
  done in $|H_{tmp}[j]| + |H|$.
  Since the $l$ points for which we compute visibility are on a grid,
  the complexity of their horizon is $|H| = O(l)$~\cite{ht:visLI},
  therefore $|H_{tmp}[j]| + |H| = O(l)$.
  In the worst case, the $l$ points are distributed over $O(\sqrt n)$
  layers, which means that the horizon $H$ will be merged $O(\sqrt n)$
  times, for a
  total of $O(l \sqrt n)$.
  %
  This is an incremental algorithm; Similar to \lores, it can be
  improved using divide-and-conquer
  to $O(l \log n)$.  
Adding the other steps,  the overall complexity of \hires is
$O(l \lg n + n/k + k \lg k + n)$.


Adding Lemma~\ref{lemma:lores} we have the following:
 \begin{theorem} 
   Given a grid terrain $T$ and a  block size $k, k > 1$, 
 a   low-resolution representation $T'$ of size $O(n/k)$  can be
 created as described above and
   the viewshed of an arbitrary point $v$ in $T$ can be computed in
   $O(n + \frac nk \lg \frac nk + k \lg k + l \cdot \lg n)$
   time, where    $l$ is the total size of blocks that are visible from $v$ in $T'$.
 \end{theorem}
 When $k$ is a constant, the running time is $O( n \lg n)$ which is no
 better than the previous upper bound~\cite{ht:visLI,kreveld:viewshed}.
 When $k>\Omega(1) $ and $l = o( n) $, the running time
 is $o( n \lg n)$ which is an improvement over the previous bound.
 In the next section we discuss the tradeoffs in choosing the block
 size $k$.

\section{Experimental Results}
\label{sec:exper}

In this section we describe experimental results showing the behavior
of our algorithm in practice and its speedup compared to previous
algorithms.

\subsection{Experimental setup} 

{\bf Algorithms tested:} We implemented our algorithm described in
Section~\ref{sec:algorithm} as a suite of three programs, \blockit,
\lores and \hires.
   \blockit computes a low-resolution grid $T'$ of a
   given grid $T$ using blocks of a given size $\sqrt k \times \sqrt
   k$.
We denote by \multires the algorithm that, given a viewpoint $v$ and a
 low-resolution grid $T'$, computes viewshed$_T(v)$ by running \lores
 followed by \hires.
When computing several viewsheds (or: the total viewshed), \blockit is
run only once and $T'$ is reused. 

 As a comparison point, we also implemented the previous algorithms
 for computing viewshed$_T(v)$: \rrr is the straightforward algorithm
 described in~\cite{franklin:sdh94} which runs in $O( n \sqrt n)$ for
 a single viewshed calculation; \visiter and \visdac are the
 horizon-based approaches from \cite{ht:visLI}, which we adapted to
 work in memory. Haverkort et al~\cite{ht:visLI} found that, in
 practice, horizons are much smaller than their theoretical worst case
 upper bound of $O(n)$, and that \visiter outperforms the recursive
 divide-and-conquer version, \visdac, even though its theoretical
 complexity is inferior. Based on these results we decided to only use
 \visiter in our experiments. For consistency the implementation of
 both \lores and \hires also use the incremental approach rather than
 divide-and-conquer.

All algorithms use the same code base, have been optimized
consistently  
and have similar interfaces. All algorithms have a flag that specifies
whether to compute a single viewshed or the total viewshed. For all
experiments described in this section we chose to compute the total
viewshed when possible (or a sampling of the total viewshed, as
described below), rather than a single viewshed. The time of a single
viewshed computation can vary significantly depending on the location
of the point. Measuring the time of the total viewshed is a more
reliable way to assess and compare viewshed algorithms, as it subsumes
the times for computing viewsheds for all points in the terrain, and
can be used as a measure for the average single viewshed running time.

\vspace{.5\baselineskip}
\noindent {\bf Viewshed sampling: } On large datasets, computing total
viewsheds takes too long and is not feasible. Therefore, we use a step
size in order to compute only a subset of the viewsheds. A step size
of $\Delta$ indicates that the algorithm only computes the viewshed
for points in both a row and column that is a multiple of $\Delta$. On
the smaller datasets (\set, \bruns and \kaweah), computing the total
viewshed takes a couple of hours so we set $\Delta=1$. For the larger
datasets we chose $\Delta$ so that the the computation finishes in a
couple of hours. For e.g. on \wash ($n = 1.06 \times 10^9$), we set
$\Delta=1000$ which means we compute approx. $1088$ viewsheds. The
step size is set consistently for all algorithms (\multires, \visiter
and \rrr).
Figure \ref{fig:viewshed_sampling} in the appendix gives details on
the step size used for each dataset and the number of  viewsheds
computed.

\vspace{.5\baselineskip}
\noindent {\bf Platform:} All modules were implemented in C and
compiled using \texttt{gcc} version 4.8.5.
The experiments were run on 
Microway Twin systems with two Intel Xeon E5-2697A v4 Broadwell 2.60
GHz Sixteen Core CPUs and 256 Gb of DDR3 2400 MHz ECC/Registered
Memory.  Each server had 16 cores however we only used one. 

 \vspace{.5\baselineskip} 
 \noindent {\bf Datasets:} We tested the algorithms on grids of
 various characteristics, described in Figure
 \ref{fig:datasets}: grid size ranges from 184,552 points (\set) to
 over $10^9$ points (\wash); some terrains are flat and some are
 mountainous; some come from Lidar and have 5m resolution or
 below ( \southport, \oahu, \clearcr), others
 come from older SRTM data at 30m resolution (\kaweah, \sierra,
 \app).


\begin{table*}
  \caption{Datasets used in the experimental analysis}
  \label{fig:datasets}
  \begin{tabular}{@{} l l r r l r l @{} }
    \toprule
    Dataset          & Grid size & $\%$ data  & True size &
    Source &  Res. & Type\\ 
    \midrule
    part of Kaweah            & $391 \times 472$
    (184,552) & 95.24\%  & $.17 \cdot 10^6$ & SRTM & 30m & mnt \\

    Brunswick, ME     & $387 \times 532$ ( 205,884)  & 100.00\% &
    $.20 \cdot 10^6$& n/a & n/a & flat \\

    Kaweah Basin        & $1,163 \times 1,424$
    (1,656,112) & 59.51\% & $.98 \cdot 10^6$ & SRTM & 30m & mnt\\
    Southport, ME     & $2,214 \times 3,674$ (8,134,236) & 87.08\% &
    $7.0 \cdot 10^6$ & Lidar & 2m & flat \\

    Sierra  Nevadas           & $2,672 \times 3,750$
    (10,020,000) & 96.13\% & $9.6 \cdot 10^6$  & SRTM &  30m & mnt\\
    Oahu, Hawai'i          & $6,685 \times 5,140$ (34, 360, 900) &
    44.89\% & $15.4 \cdot 10^6$ & Lidar & 5m &mnt\\
    Appalachian Mnt & $8,479 \times  7,850$ (66,560,150) &
    51.39\% & $34.2 \cdot 10^6$ & SRTM & 30m &mnt\\
    Clear Creek, IA & $13,040 \times 8,749$ (114,086,960) & 98.73\%
    & $ 112.6 \cdot 10^6$ & Lidar & 1m & flat\\ 
    Portland , ME      & $11,989 \times 10,693$ ( 128,198,377) & 100.00\%
    & $122.2 \cdot 10^6$ & Lidar & 10m & flat \\
    Puget Sound, Wash. & $31,866 \times 33,454$ (1,066,045,164)
    & 99.98\% & $1,065.8 \cdot 10^6$& SRTM & 10m    & flat  \\ 
\bottomrule
    \bottomrule  
\end{tabular}
\end{table*}

\subsection{Effect of block size}

The first question we investigated is the dependency of \lores and
\hires on the block size $k$, and what value, or range of values, of
the block size results in optimal overall running time of \multires.
To this end, for each dataset, we ran \multires with various values of
$k$, ranging from $k = 3 \times 3$ to $k= 20\times 20$.

\begin{figure*}[t]
  \centering
  \begin{subfigure}{.4\textwidth}
    \centering
    \includegraphics[width=\textwidth]{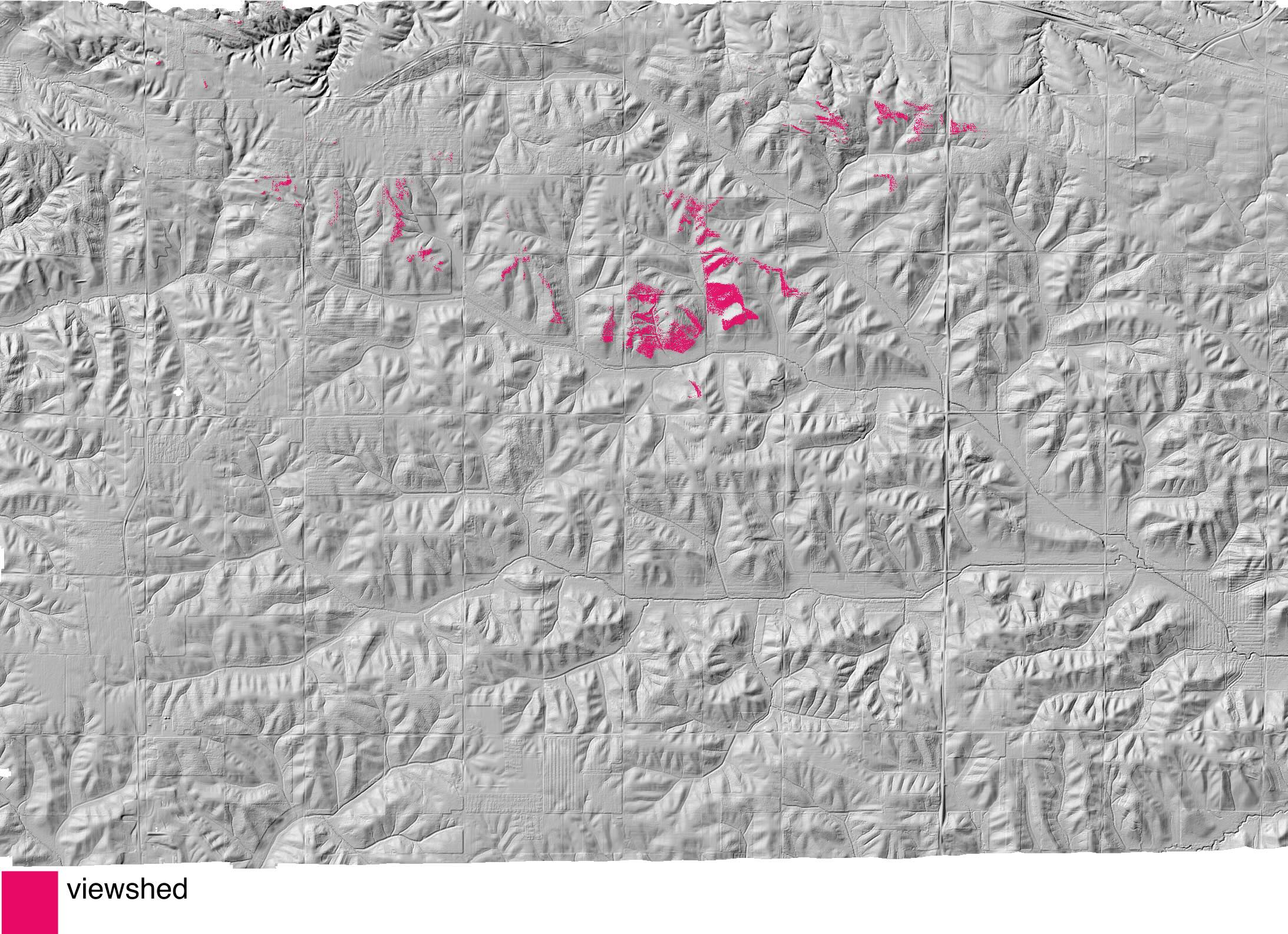}
    \caption{The viewshed}
  \end{subfigure}
  \begin{subfigure}{.4\textwidth}
    \centering
    \includegraphics[width=\textwidth]{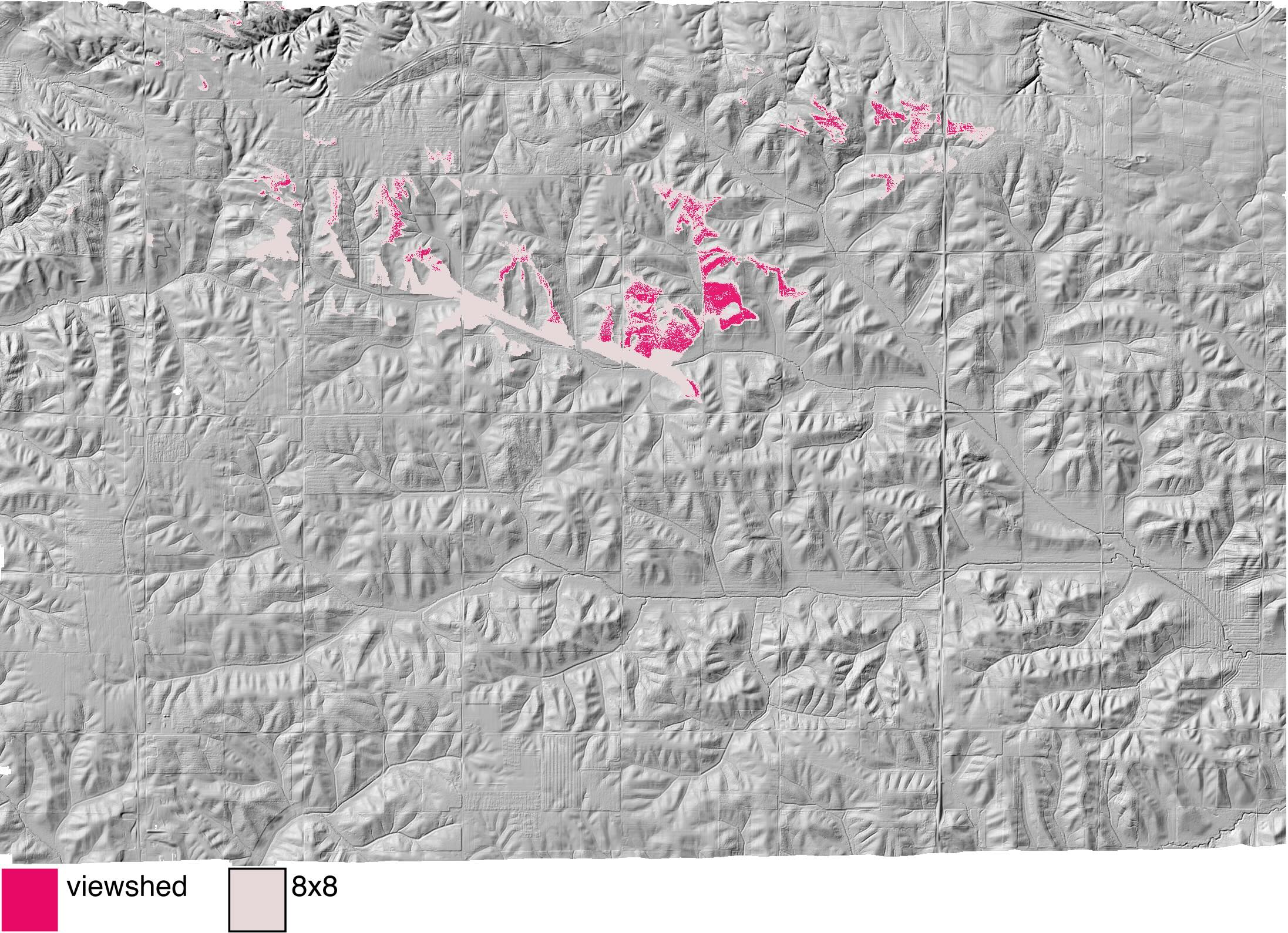}
    \caption{\lores viewshed, $k=8\times8$}
  \end{subfigure}
  \begin{subfigure}{.4\textwidth}
    \centering
    \includegraphics[width=\textwidth]{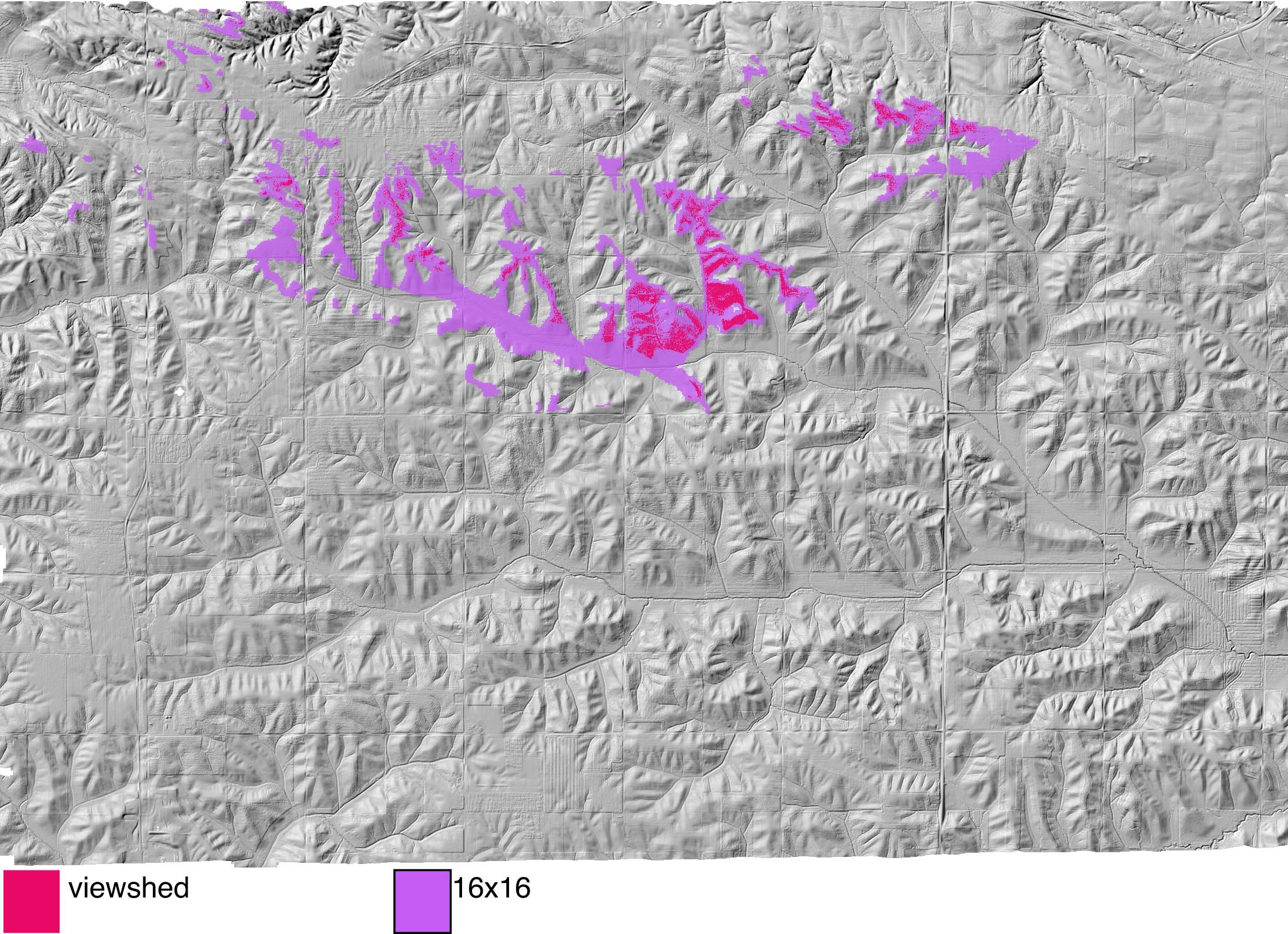}
    \caption{\lores viewshed, $k=16\times16$}
  \end{subfigure}
  \begin{subfigure}{.4\textwidth}
    \centering
    \includegraphics[width=\textwidth]{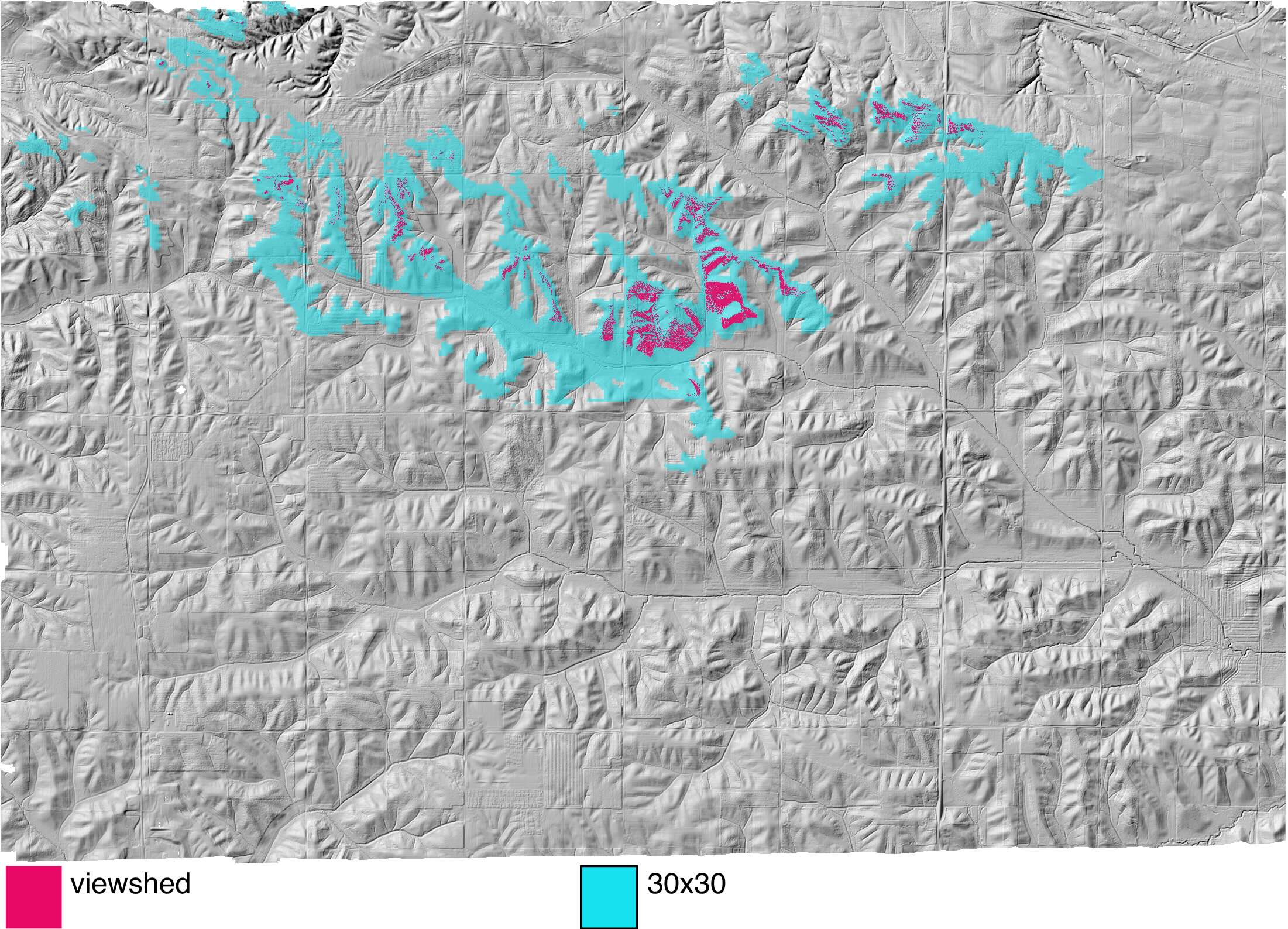}
    \caption{\lores viewshed,  $k=30\times30$}
  \end{subfigure}
  \begin{subfigure}{.4\textwidth}
    \centering
    \includegraphics[width=\textwidth]{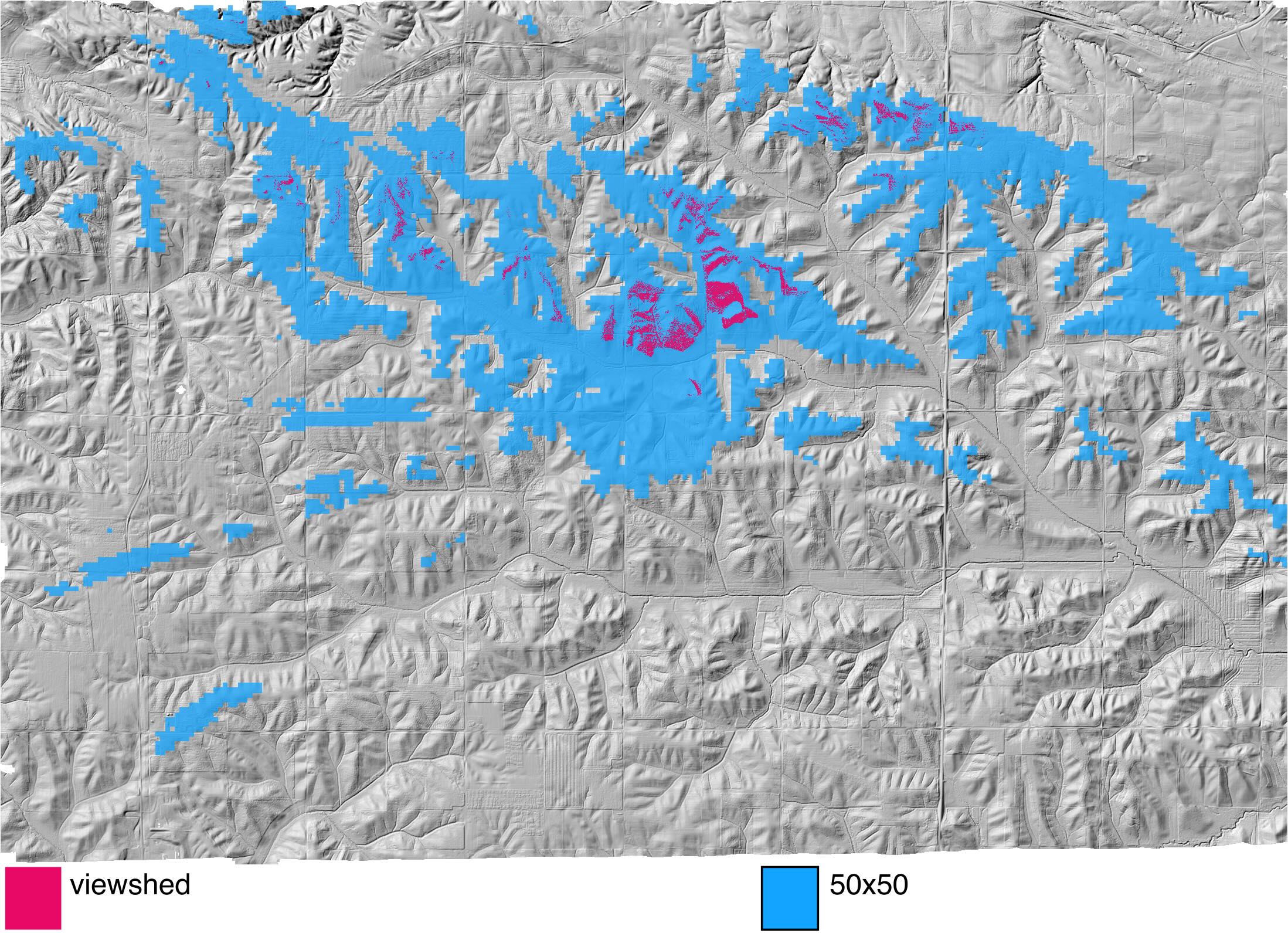}
    \caption{\lores viewshed,  $k=50\times50$}
  \end{subfigure}
  \begin{subfigure}{.4\textwidth}
    \centering
    \includegraphics[width=\textwidth]{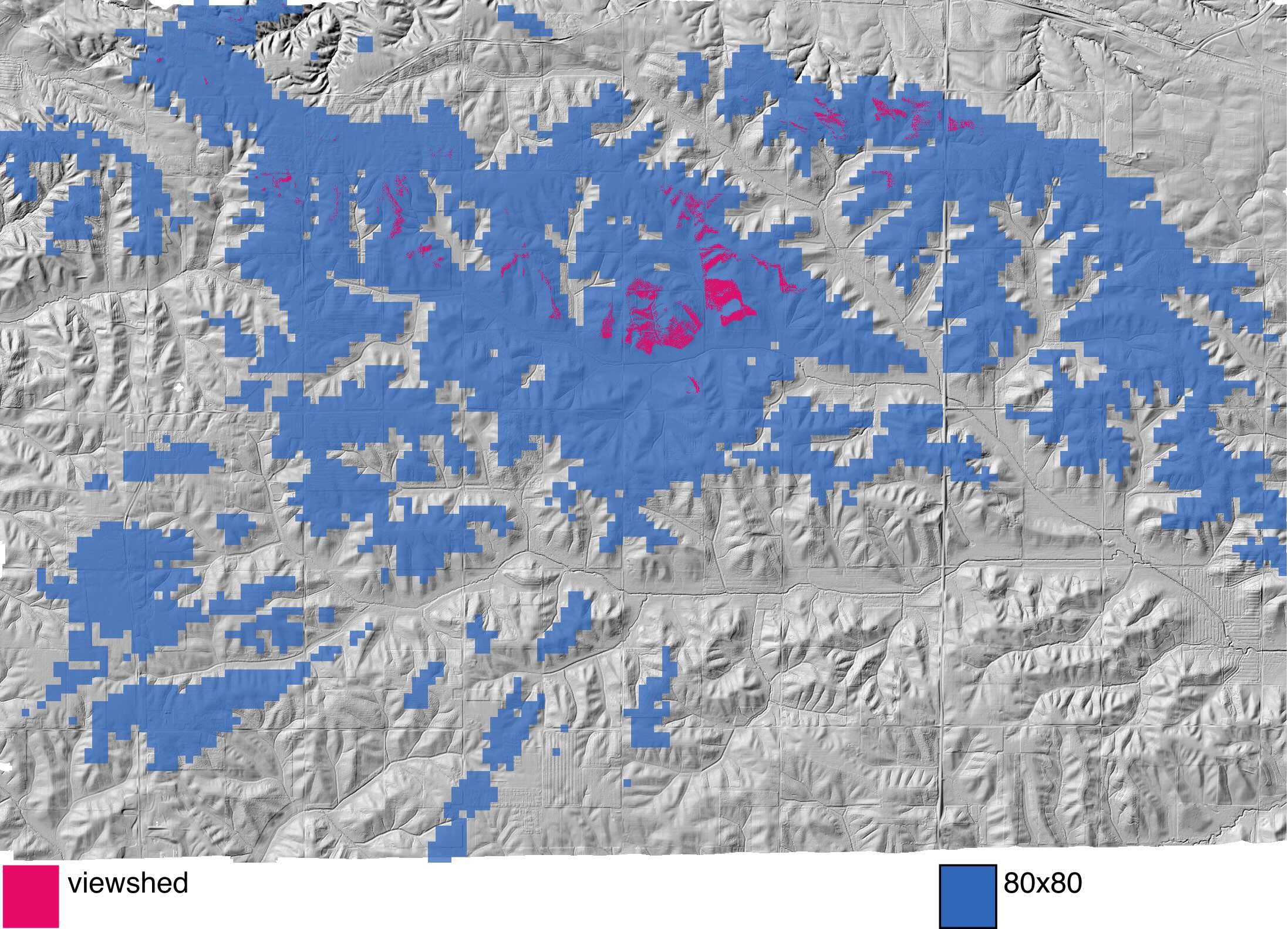}
    \caption{\lores viewshed, $k=80\times80$}
  \end{subfigure}
  \begin{subfigure}{.4\textwidth}
    \centering
    \includegraphics[width=\textwidth]{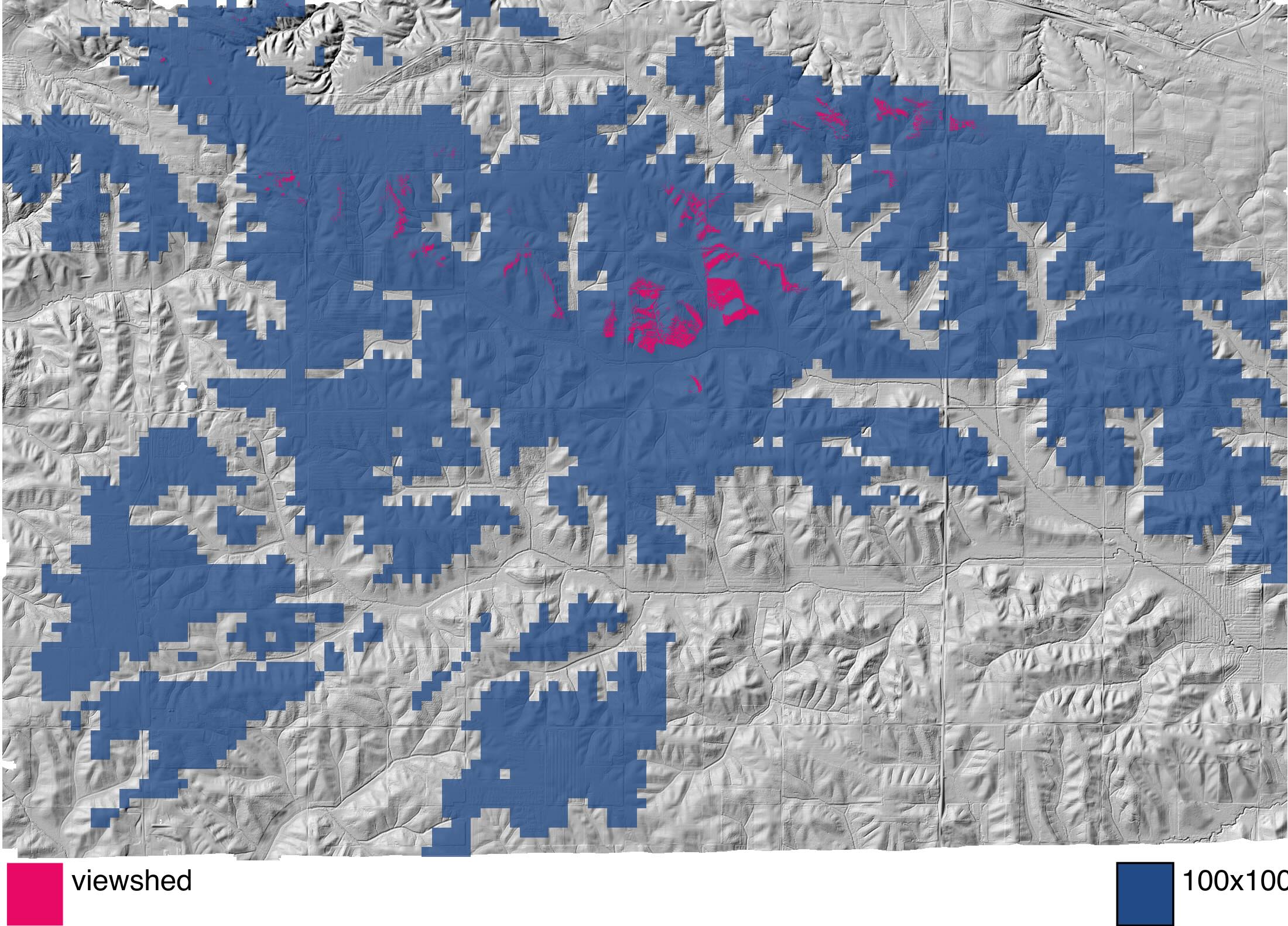}
    \caption{\lores viewshed, $k=100\times 100$}
  \end{subfigure}
  \begin{subfigure}{.4\textwidth}
    \centering
    \includegraphics[width=\textwidth]{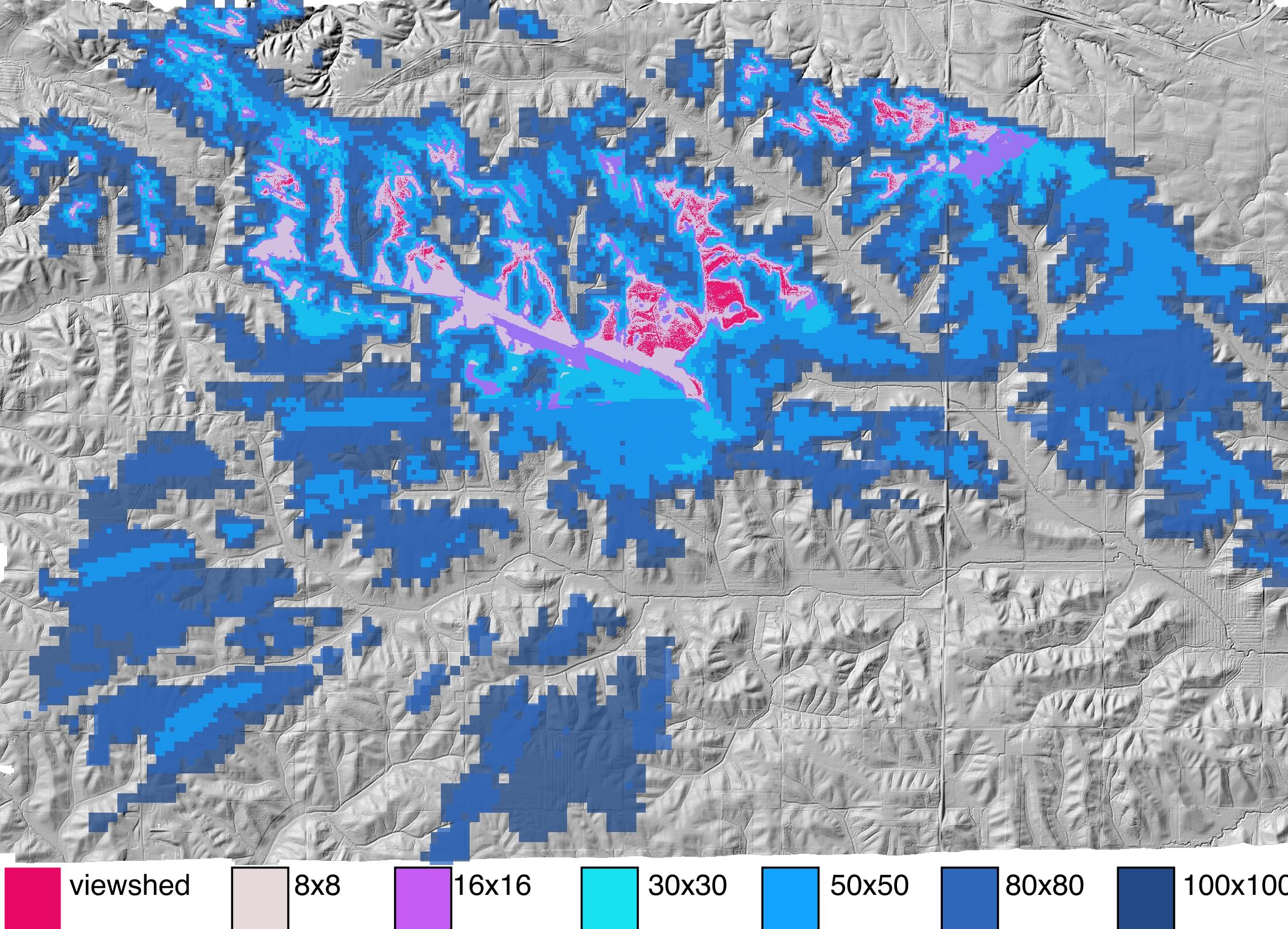}
    \caption{Overall: Effect of block size}
  \end{subfigure}
  
  \caption{\clearcr dataset: The viewshed overlaid on the \lores
    viewsheds computed with various block
    sizes for viewpoint at     $(r=4,000, c=7,000)$. A close-up on   central area is provided in the appendix. 
  }
\label{fig:clear_creek_blocksize}
\end{figure*}

Visual results for \clearcr are shown in
Figure~\ref{fig:clear_creek_blocksize} for the entire dataset, and in
Figure~\ref{fig:clear_creek_zoom} (appendix) for a zoomed in view, and
Note that the high-resolution viewshed is always a subset of the
low-resolution viewshed, as stated by the algorithm's invariant. We
see that when the block size is small, the low-resolution viewshed
matches the high-resolution viewshed closely, and, as $k$ increases,
the low-resolution viewshed gets larger. Recall that we insert each
block $B$ in the horizon as a segment from $\Theta_1$ to $\Theta_2$,
where $[\Theta_1, \Theta_2]$ is the span of $B$, at zenith equal to
$z_{min}(B)$; this represents a lower bound for the horizon of $B$. A
block is guaranteed to be invisible if a segment with span $[\Theta_1,
\Theta_2]$ at zenith equal to $z_{max}(B)$ is completely below the
horizon. Therefore, the quality of the low-resolution viewshed depends
on how close the minimum and maximum zenith of a block are to the true
horizon of the block. For a small block size or flat areas, the
minimum and maximum elevation in a block are close in value. As $k$
increases, the difference between the minimum and maximum elevation in
a block increases, which will lead to fewer blocks being marked as
invisible, or a larger low-resolution viewshed.

\begin{table*}[ht]
  \caption{Clear Creek ($n=114.1 \cdot 10^6$) results: Running time (in seconds) of
    \multires; the speedup relative to \visiter for various block
    sizes; \lores and \hires viewshed sizes (as a percentage of
    grid size).}
  \label{table:iowa_time}
  \begin{tabular}
    {@{} r c r r r r r r c @{}}
    \toprule
    $ k$  & \blockit& \lores & \hires& \multires  &
    Speedup & \lores & \hires & \lores /\hires  \\ 
    & time & time & time & time && VS size & VS size & VS size \\ 
    \midrule
    $8\times 8$  & 4.95      & 3,503.32 & 3,778.50 & 7,281.82 & 20.93   & 1.22\%  & 0.14\%  & 8.87   \\
    $9\times 9$  & 4.27      & 2,661.75 & 3,947.44 & 6,609.19 & 23.06   & 1.33\%  & 0.14\%  & 9.72 \\
    $10\times 10$ & 3.93      & 2,015.36 & 4,034.10 & 6,049.46 & 25.19    & 1.45\%  & 0.14\%  & 10.61 \\
    $11\times 11$ & 3.59      & 1,640.72 & 4,134.99 & 5,775.71 & 26.39   & 1.57\%  & 0.14\%  & 11.41    \\
    $12\times 12$ & 3.32      & 1,342.05 & 4,179.93 & 5,521.98 & 27.60    & 1.68\%  & 0.14\%  & 12.29    \\
    $13\times 13$ & 3.12      & 1,100.67 & 4,416.26 & 5,516.93 & 27.62   & 1.80\%  & 0.14\%  & 13.10  \\
    $14\times 14$ & 2.96      & 928.89   & 4,512.18 & 5,441.07 & 28.01  & 1.91\%  & 0.14\%  & 13.96           \\ 
    $15\times 15$ & 2.85      & 802.45   & 4,652.96 & 5,455.41 & 27.94    & 2.04\%  & 0.14\%  & 14.84 \\
    $16\times 16$ & 2.74      & 697.88   & 4,786.52 & 5,484.40 & 27.79   & 2.15\%  & 0.14\%  & 15.65 \\
    $17\times 17$ & 2.62      & 613.49   & 4,843.34 & 5,456.83 & 27.93   & 2.27\%  & 0.14\%  & 16.53  \\
    $18\times 18$ & 2.58      & 541.99   & 5,023.91 & 5,565.90 & 27.38  & 2.38\%  & 0.14\%  & 17.41        \\
    $19\times 19$ & 2.49      & 485.57   & 5,153.81 & 5,639.38 & 27.02   & 2.50\%  & 0.14\%  & 18.30    \\
    $20\times 20$ & 2.40      & 435.79   & 5,272.45 & 5,708.24 & 26.70    & 2.63\%  & 0.14\%  & 19.19 \\ \bottomrule
  \end{tabular}
\end{table*}

We give detailed numbers and running times for the same dataset,
\clearcr ($n=112.6 \cdot 10^6$) in Table~\ref{table:iowa_time}; the
results for the other datasets are very similar and are omitted.
%
On \clearcr the step size is set to $\Delta=100$ which means a total
of $11,348$ viewsheds are computed.
The average viewshed size on \clearcr is $0.14\%$ of the grid size,
where the average is taken over all the computed viewsheds. We
consider the ratio of the average size of the low-resolution viewshed
as computed by \lores, to the average viewshed size as computed by
\hires (sometimes we refer to the viewshed computed by \hires as the
high-resolution viewshed). At $k= 8 \times 8$ the average
low-resolution viewshed size is 1.22\% of the grid size, which is a
factor of 8.87 times larger than the average viewshed size. As $k$
increases, the size of the low resolution viewshed goes up; at $k=20
\times 20$ the low-resolution viewshed size is 2.63\%, which is a
factor of 19.19 times larger than the average viewshed size.

\begin{figure*}[t]
  \begin{center} 
    \begin{subfigure}{.45\textwidth} 
      \centering{
        \includegraphics[width=.85\linewidth]{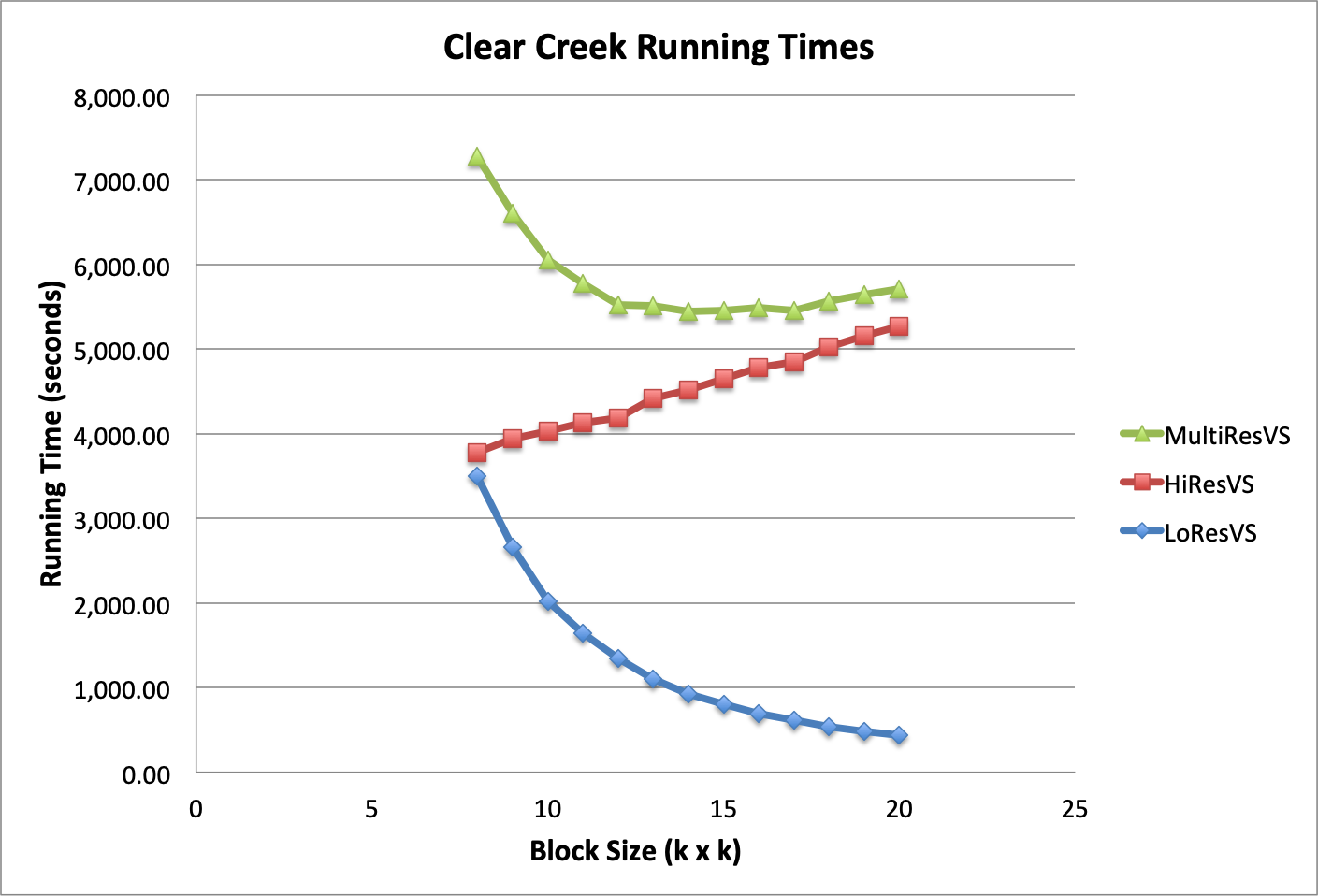}
        \caption{Running time of \multires}
      }
      \end{subfigure} 
 \begin{subfigure}{.45\textwidth} 
   \centering{
     \includegraphics[width=.85\linewidth]{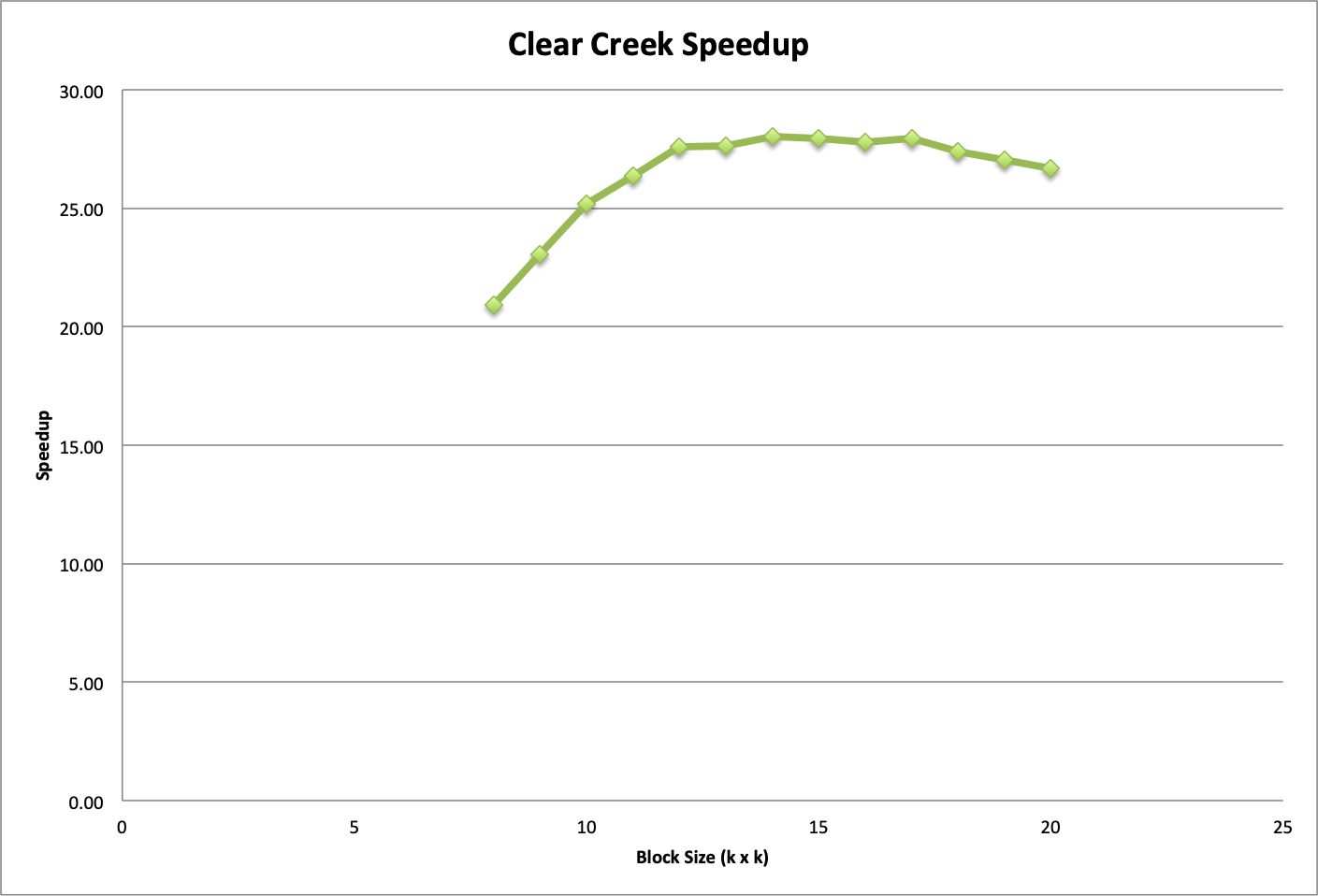}
     \caption{Speedup of \multires over \visiter}
   }
 \end{subfigure} 
\end{center}
\caption{ Clear Creek results: Running time and speedup with
  various block sizes. }
\label{fig:clearcreek}
\end{figure*} 

Figure~\ref{fig:clearcreek} shows the running time of \lores and
\hires and the speedup of \multires over \visiter on \clearcr, for
various block sizes.
We see that the running time of \lores goes down as the block size $k$
increases; this is to be expected, as the complexity of \lores is
$O(\frac nk \lg \frac nk)$. As $k$ increases, \lores is faster, but is
less able to filter out blocks that are guaranteed to be invisible,
and the size $l$ of the low-resolution viewshed goes up. As a result,
the running time of \hires goes up. Again, this is to be expected as
the complexity of \hires is $O(l \lg n + n/k + k \lg k + n)$, and with
$k << n$ and ignoring the $O(n)$ term, the dominant term is $O(l \lg
n)$.

These two opposite trends, \lores getting faster and \hires getting
slower as $k$ increases, are optimized on \clearcr at a block size
of $k=14\times 14$. At this optimal block size, \lores runs in 928.9
seconds and labels $1.91\%$ of the grid, or $l=.0191n=2.1\cdot 10^6$
points, as potentially visible; \hires runs in additional 4,512.2
seconds to refine the visibility of these points. Overall, at $k=14
\times 14$, the low-resolution viewshed size is a factor of 13.96
larger than the viewshed size, and \multires is a factor of 28.01
times faster than \visiter. The speedup is plotted in
Figure~\ref{fig:clearcreek}(b). The peak is at at $k=14 \times 14$,
however we note that all values of $k$ between $10 \times 10$ and $20
\times 20$ achieve a speedup of at least 25 over \visiter.

\subsection{Optimal Block Size}

\begin{figure}[t]
\centering
\includegraphics[width=0.6\linewidth]{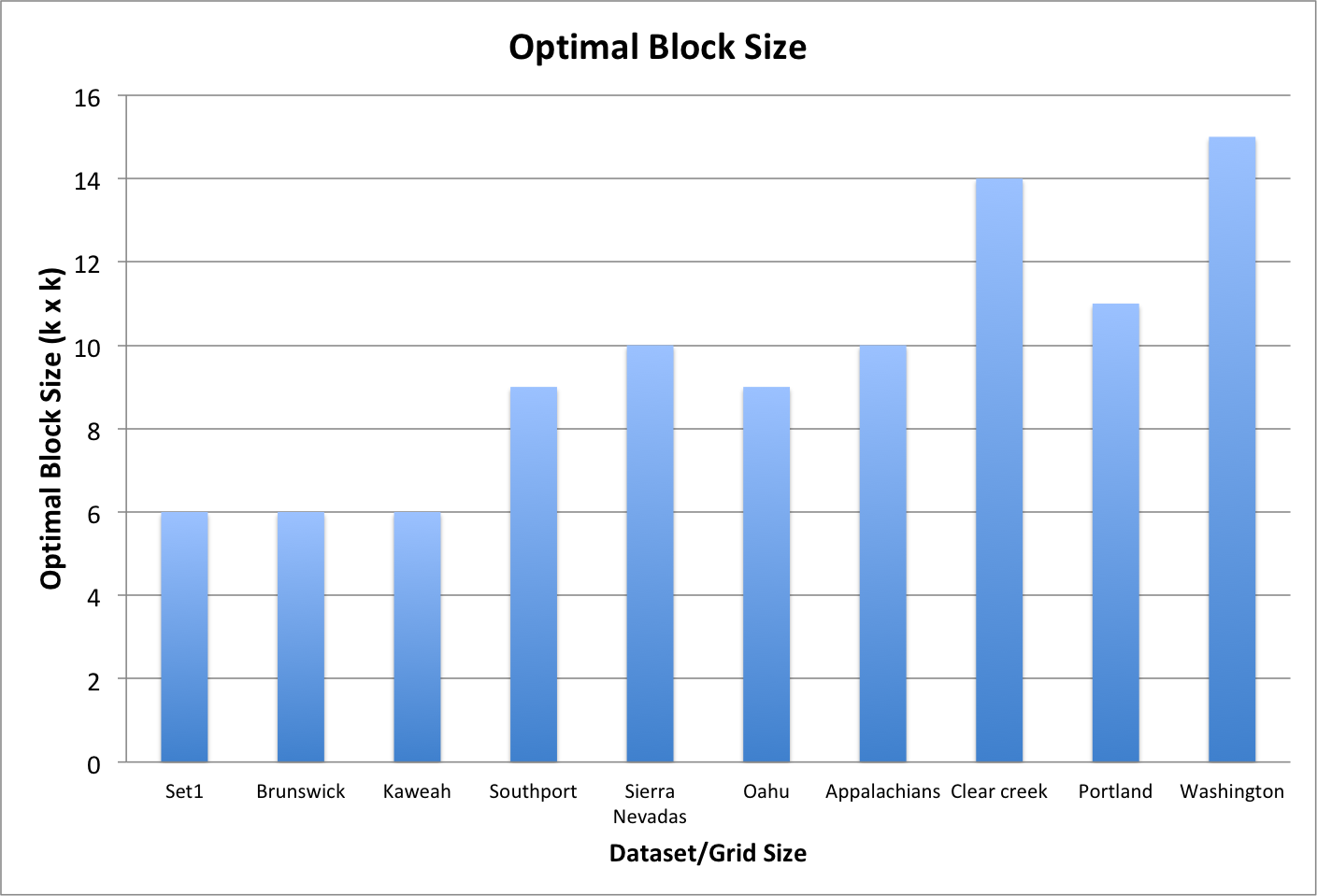}
\caption{Optimal block size for \multires by dataset}
\label{fig:opt_block_size}
\end{figure}

For each dataset, we denote the \emph{optimal block size} the value
$k$ that results in overall minimum running time for \multires. Since
the running time of \multires depends on the size of the
low-resolution viewshed, which, in turn, depends on the terrain
topography, the optimal block size can only be determined empirically.
For each dataset we ran experiments with various values of
$k$, ranging from $k = 3 \times 3$ to $k= 20\times 20$.
The results are summarized in Figure~\ref{fig:opt_block_size} which
shows the optimal block size per dataset.

The optimal block size starts at $k=6 \times 6$ for the smaller
datasets (\set, \bruns, \kaweah); it goes up to $k= 10 \times 10$ for
the medium datasets (\southport, \sierra, \oahu, \app, \portland), and
further up to $k=15 \times 15$ for \clearcr, and \wash. We see that
the optimal block size increases with grid size $n$. A likely
explanation is that as the grid size increases, the percentage of grid
points contained in a block of size $k$ decreases,
 and therefore, the size of the invisible region relative to
 the block size $k$ increases, which means that the number blocks of
 size $k$ needed to cover the invisible region increases.
 \lores will be able to filter out the same region using a
 larger block size, and the drop in running time due to the increased
 $k$
 will be significant. For example, on \clearcr (refer
 toTable~\ref{table:iowa_time}): with a block size of $8
 \times 8$, \lores takes $3,503$ seconds to label $l=.0122n$ as
 potentially visible, while with $k = 14 \times 14$ \lores takes $928$
 seconds to label $l=.0191n$ as potentially visible. The increase in
 $l$ is small, while the decrease in \lores time is significant,
 which will make a larger $k$ overall more efficient. We note that for
 the smaller  datasets $l$ increases much faster with $k$.

An additional factor for the increase of the optimal block size with
$n$ is that the average viewshed size on the datasets in our
experiments seems to be decreasing slightly as the grid size increases
(see Table~\ref{table:avg_vs_size}; \wash is an exception likely due
to Puget Sound being a relatively flat area with lots of water, which
leads to larger viewsheds). A smaller viewshed means a larger
invisible area,
so \lores
is  able to filter out the same region using a larger block size. 

The optimal block size is consistent across datasets of similar size,
for e.g. \southport, \sierra, \oahu, which have similar size, all have
optimal $k= 9 \times 9$ or $k=10 \times 10$. However optimal block
size shows dependence on other terrain characteristics (topography);
for e.g. \portland~is larger than \clearcr, but the optimal block size
for \portland~is $k=11\times11$ compared to $k=14\times 14$ for
\clearcr; similarly, \oahu~is larger than \sierra, but its optimal
block size is $k=9 \times 9$ as opposed to $k=10 \times 10$ for
\sierra.

For any given dataset, the dependence of \multires running time on
block size is fairly stable, with optimal performance achieved at a
range of block values. For e.g. on \clearcr (shown in
Table~\ref{table:iowa_time}), block sizes of $k=14 \times 14$ through
$k=17 \times 17$ have a similar speedup. This behaviour is consistent
on all datasets:
on \portland, block sizes $k=11 \times 11$ and $k=12 \times 12$ give
similar speedup, and on \wash, blocks sizes of $k=15 \times 15$
through $k=20 \times 20$ all give a similar speedup.


\begin{table}[t]
\caption{Average  viewshed size of \lores and \hires as a percentage of 
  grid size  at optimal block size}
\label{table:avg_vs_size}
\centering
\begin{tabular}{@{}llll@{}}
\toprule
Dataset      & \lores & \hires & \lores/\hires \\ \midrule
set1         & 17.76\%          & 3.40\%           & 5.22        \\
brunswick    & 18.76\%          & 2.38\%           & 7.88        \\
kaweah       & 10.83\%          & 2.35\%           & 4.61        \\
southport    & 7.21\%           & 0.97\%           & 7.43        \\
sierra       & 4.94\%           & 0.81\%           & 6.08        \\
oahu         & 8.85\%           & 1.88\%           & 4.71        \\
appalachians & 3.40\%           & 0.08\%           & 40.19       \\
clear creek  & 1.91\%           & 0.14\%           & 13.96       \\
portland     & 2.87\%           & 0.61\%           & 4.72        \\
washington   & 3.00\%           & 2.29\%           & 1.31        \\ \bottomrule
\end{tabular}
\end{table}

Since \multires is an output-sensitive algorithm whose theoretical
complexity depends on the size of the viewshed, understanding how the
size of the viewshed behaves on real datasets, what are the factors
that affect it, and the relation between \lores viewshed size,
(\hires) viewshed size and block size $k$, are all important factors
for understanding how \multires performs in practice.
Table~\ref{table:avg_vs_size} shows, for each dataset, the
\emph{average} viewshed size, and the \emph{average} \lores viewshed
size, $l$, when using the optimal block size for that particular
dataset. The averages were calculated for each dataset over all
viewsheds that were computed part of the total viewshed calculation
(recall that our total viewshed algorithm uses a step $\Delta$ to
control how many viewshed calculations to perform; see
Table~\ref{fig:viewshed_sampling} in the appendix).
First, we note that the average viewshed size is less than 3.4\% of
grid size ($.034n$) for all datasets, with 6 out of the 10 datasets
having an average viewshed size below $1\%$. It is not surprising that
mountainous regions have small viewsheds (e.g. \sierra, \kaweah,
\app), but we were surprised to find that relatively flat terrains
like \portland, \clearcr and \wash have small viewsheds as well. The
datasets \southport, \oahu, \portland, and \clearcr are all derived
form high-resolution Lidar data (typically Lidar data has 15cm
vertical accuracy), and their average viewshed size is $0.97\%,
1.88\%, 0.14\%$ and $0.61\%$, respectively. Thus it seems that
Lidar-derived digital elevation models lead to small viewsheds even
for relatively flat topographies.

The average \lores viewshed size (with $k$ set to its optimal value )
is just below 20\% for the smaller datasets, and drops to approx. 3\%
for the larger datasets. We see a clear negative correlation between
the size of the dataset and the average low-resolution viewshed size.
This is to be expected, as the optimal block size does not increase
proportionally to the grid size, so on larger grids, the percentage of
points contained in the optimal block size is much smaller. Therefore,
on large grids, \lores marks a much greater percentage of points as
invisible.

Table~\ref{table:avg_vs_size} also gives the ratio of the average
viewshed size of \lores (run with optimal block size), and the average
viewshed size of \hires. We do not see a clear trend. The ratios for
most of the datasets are fairly similar with the average
low-resolution viewshed size about 5 to 7 times larger than the
average viewshed size. However, the ratio for \wash~is much lower than
the others, while the ratio for \app~(and also \clearcr) is much
higher. 
Interestingly (and puzzlingly), these three datasets with the largest
ratio of \lores to \hires size are precisely those where the speedup
of \multires is largest ---- see below.


\subsection{Speedup over  previous algorithms}

\begin{figure}[t]
  \centering
  \includegraphics[width=0.6\linewidth]{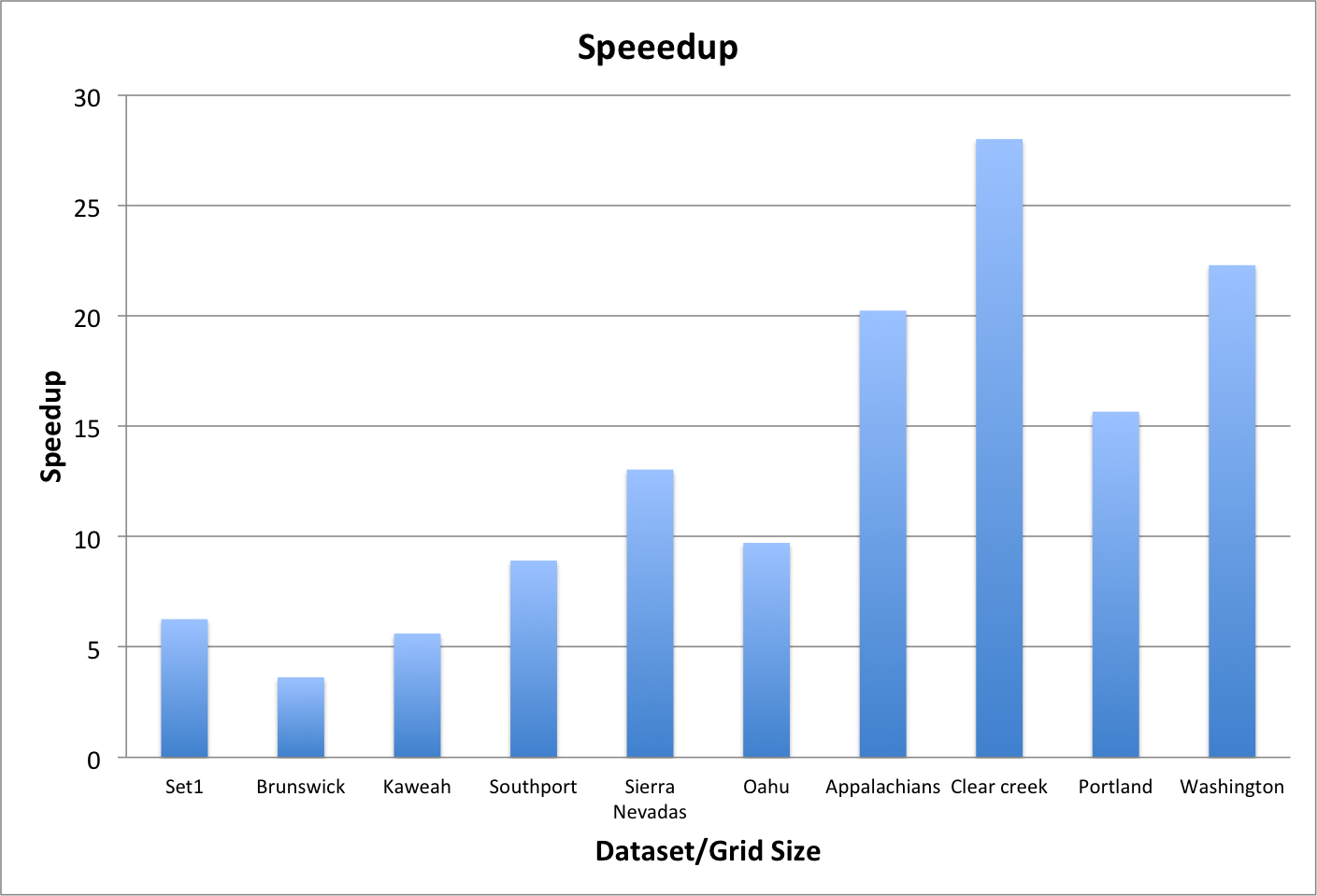}
  \caption{Speedup of \multires over \visiter  by dataset}
  \label{fig:speedup}
  \end{figure} 

We compared \multires with the two previous algorithms that compute
viewshed$_T(v)$: naive (or \rrr~\cite{franklin:sdh94}) and \visiter
~\cite{ht:visLI}.
%
First, we ran the naive algorithm on the three smallest datasets,
\set, \bruns, and \kaweah;  on all these datasets the  step size is
$\Delta=1$ so the computation is a true total viewshed. 
%
We found that the naive algorithm's running time is a factor of 2.25,
2.11 and 5.03, respectively, slower than \visiter. Due to the slow
running time of the naive algorithm, we have not run experiments on
any of the larger datasets, and we only compared \multires and
\visiter.

The speedup of \multires to \visiter for all datasets is shown in
Figure~\ref{fig:speedup}. First, we see that the speedup increases
with grid size, starting at 6.26 for \set~and reaching 22.30 for
\wash. This is because the optimal block size increases with grid
size, and the low-resolution viewshed size at optimal block size
decreases with grid size (Table~\ref{table:avg_vs_size});
Therefore, on larger grids, \lores is able to mark a greater
percentage of the points as invisible and the speedup is larger.

  Overall the speedup increases with dataset size, but depends on
  other dataset characteristics besides
  size. For example, the speedup on \portland, which is flat,  is
  15.66, while the speedup for \newline \app, which is smaller but
  mountainous, is 20.24. In general, we expected the speedup to be
  smaller on flat terrains because flat terrains have larger
  viewsheds.
  Note however that the speedup is significant across all
  datasets, even for flat terrains. In particular, \bruns, 
  \portland  and \wash, which are flat, see speedups of 3.63, 
  15.66, and 22.30, respectively.

  The best speedup is achieved on \clearcr, which is the dataset
  with the best horizontal resolution. It is overall flat, but we
  suspect has a lot of small variations locally. The need for faster
  total viewshed algorithms is born out
  of the availability of large high-resolution datasets and the need
  to model terrains with a high level of accuracy, so the fact that
  our algorithm achieves the highest speedup on this type of data is
  promising.

\section{Conclusion and Future Work}
\label{sec:concl}

This paper describes a novel approach for computing viewsheds based on
the idea of using a lower-resolution version of the grid to quickly
identify blocks of points that are guaranteed to be invisible. In
practice, our algorithm achieves a speedup of 20 or more on large,
high-resolution grids compared to previous algorithms.

Our algorithm does not introduce approximation and computes the
ground-truth viewshed. However the approach can be used to speed up
\emph{any} viewshed algorithm, including approximate ones: run \lores
to identify blocks that are guaranteed to be invisible, then use any
(approximate) viewshed algorithm to refine the low-resolution
viewshed. If the viewshed algorithm runs in $T(n)$, this approach will
run in $O(n/k \lg n/k + T(l)) < T(n)$. Approximation can be introduced
in \lores as well by
using a different approach to compute the low-resolution viewshed that
runs faster than $O(n/k \lg n/k)$, or by using smaller upper bound or larger
lower bound for $z_{max}(B)$ and $z_{min}(B)$, respectively.

As direction for future work, the next step is to extend the approach
from two levels to a hierarchy of levels. One idea is to use small
blocks around the viewpoint and larger blocks further away from the
viewpoint. Another idea is to use a quadtree-like approach: start with
a low-resolution representation and determine if a block is
potentially visible; if yes, divide it into four smaller blocks and
repeat. This way a region in the terrain will dynamically adjust its
block size as needed during the computation. A hierarchy of
low-resolution grids can be pre-computed and if their sizes are chosen
as a geometrically decreasing series the overall space is $O(n)$.

\paragraph{Acknowledgements} I would like to thank Herman Haverkort
for his insights and contribution in the early stages of this project,
and to Andrew Prescott for taking on this project as his honors thesis
and transforming it into efficient code.

%
\newpage
\appendix

 \begin{figure*}[h]
 \centering
 \includegraphics[scale=0.35]{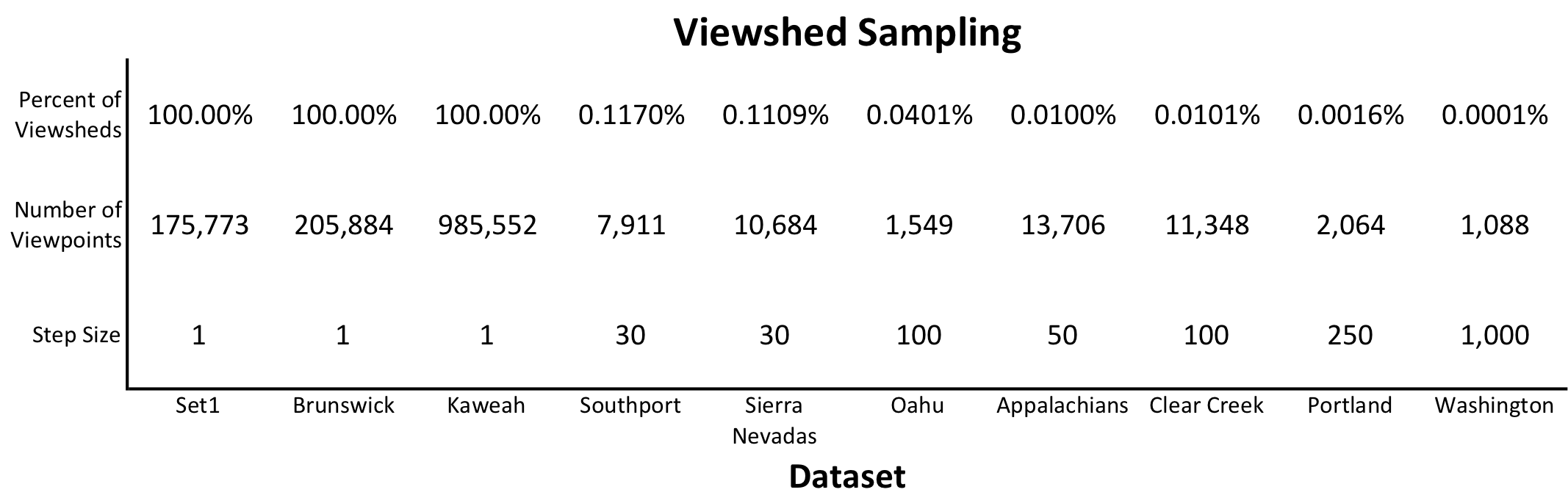}
 \caption{Step size, number of viewpoints and percent of viewsheds used for each dataset}
 \label{fig:viewshed_sampling}
 \end{figure*}



\begin{figure*}[h]
 \begin{subfigure}{.45\textwidth}
   \centering
   \includegraphics[width=0.9\linewidth]{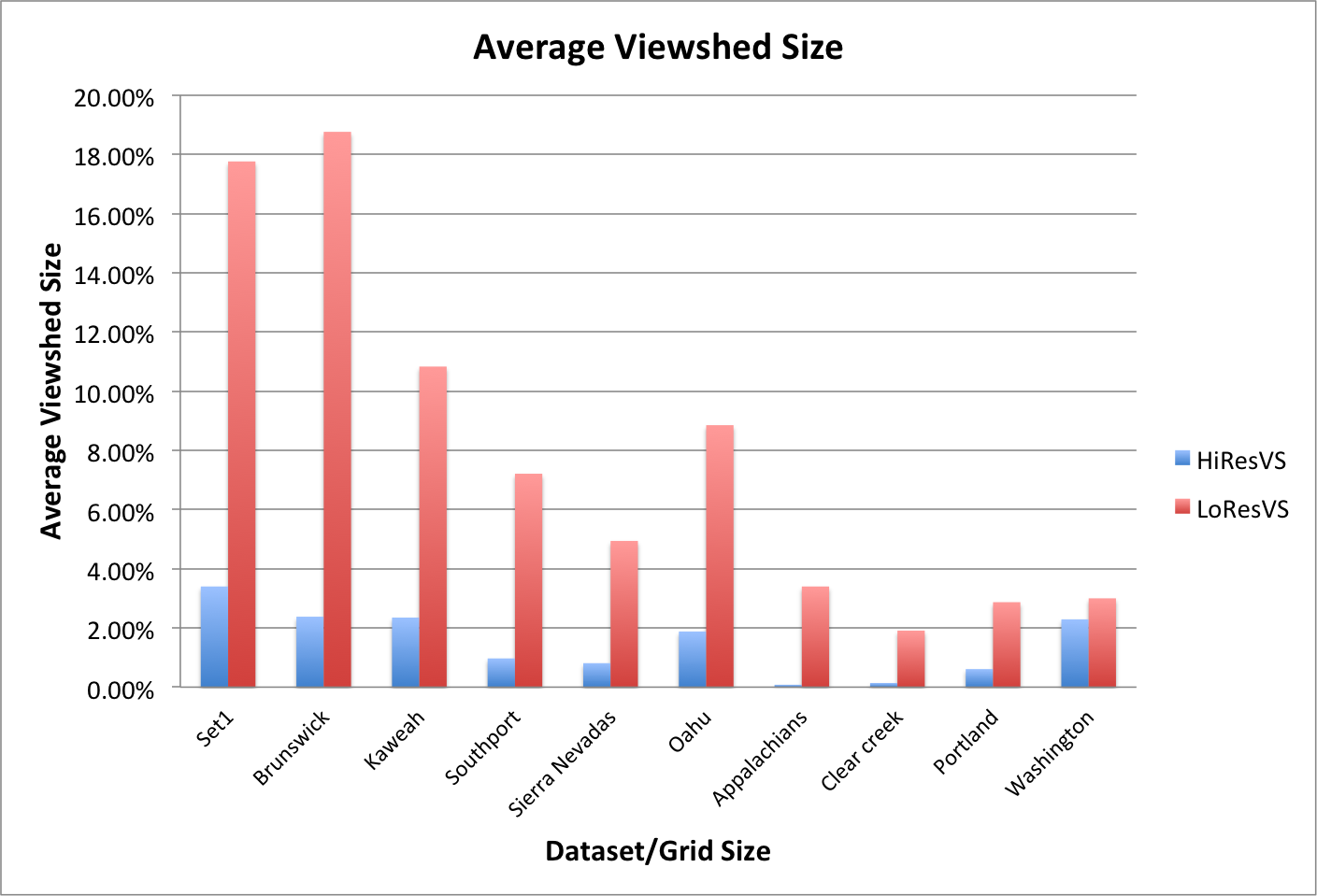}
   \caption{Avg. viewshed size of \lores and \hires by dataset}
 \end{subfigure} 
 \begin{subfigure}{.45\textwidth}
   \centering
   \includegraphics[width=0.9\linewidth]{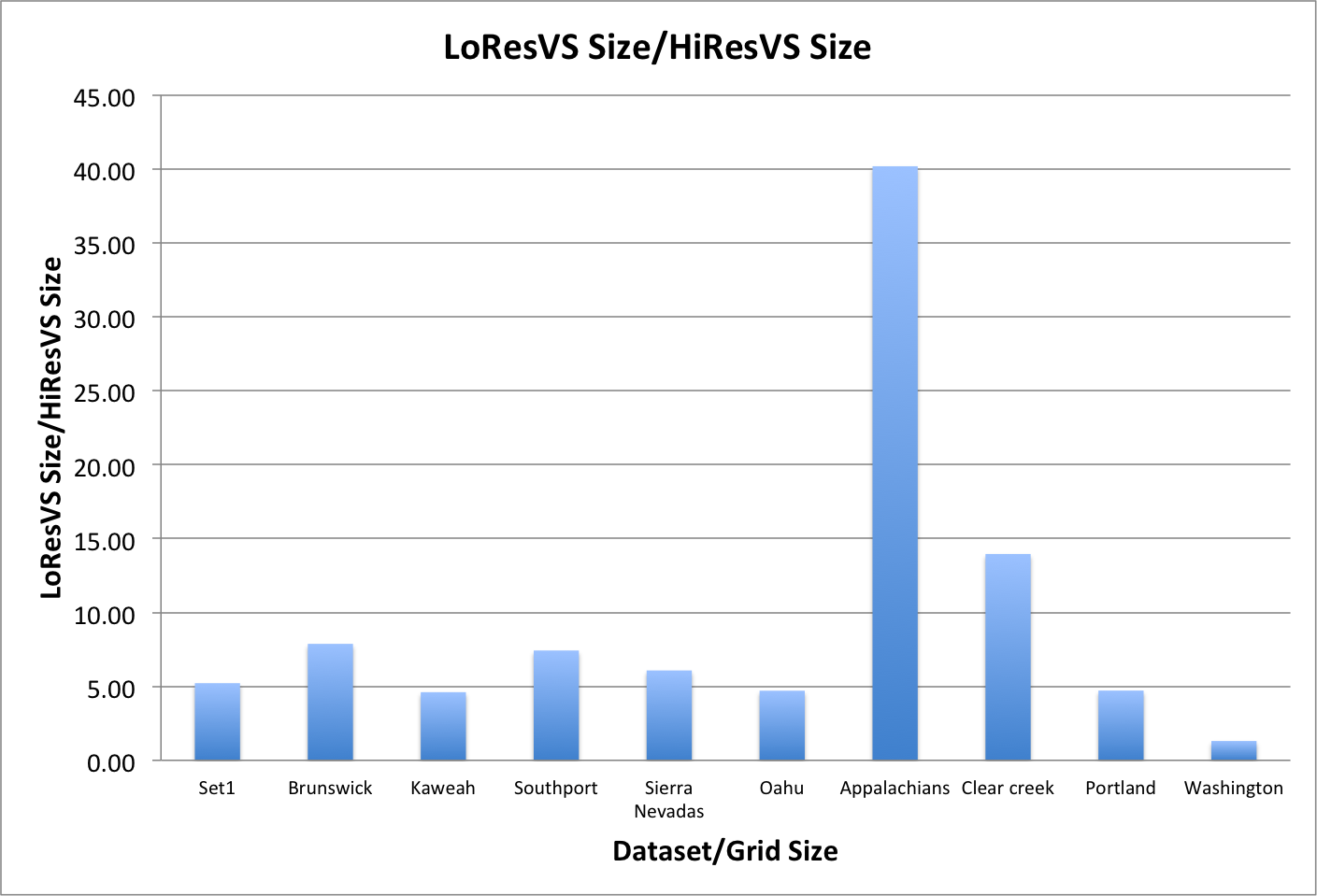}
\caption{Avg. viewshed size ratio of \lores to \hires by dataset}
\end{subfigure} 
\caption{Viewshed size by dataset}
\label{fig:avg_vs_size_ratio}
\end{figure*}


\begin{figure*}[t]
  \centering
  \begin{subfigure}{.4\textwidth}
    \centering
     \includegraphics[width=\textwidth]{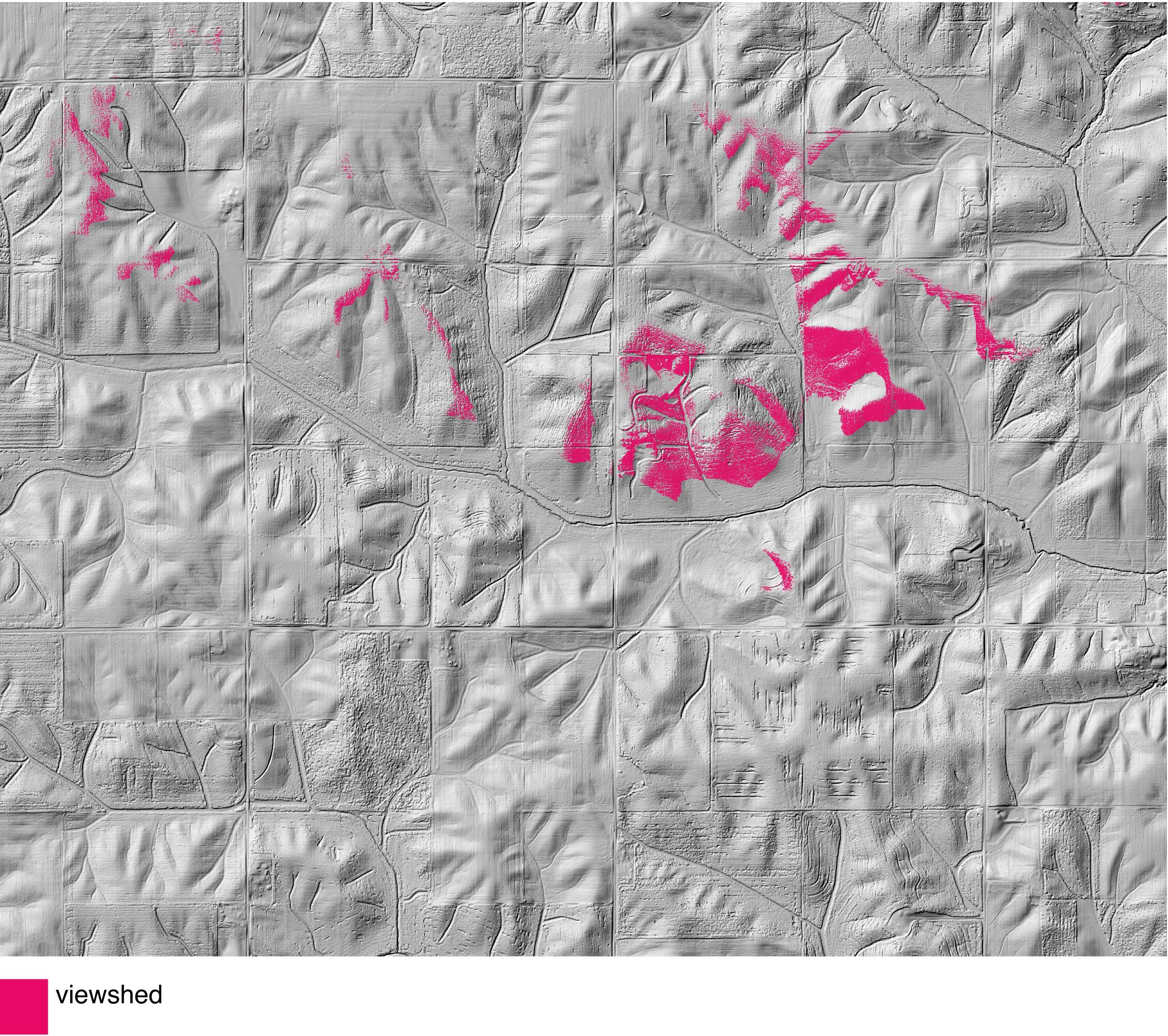}
    \caption{The viewshed}
  \end{subfigure}
  \begin{subfigure}{.4\textwidth}
    \centering
     \includegraphics[width=\textwidth]{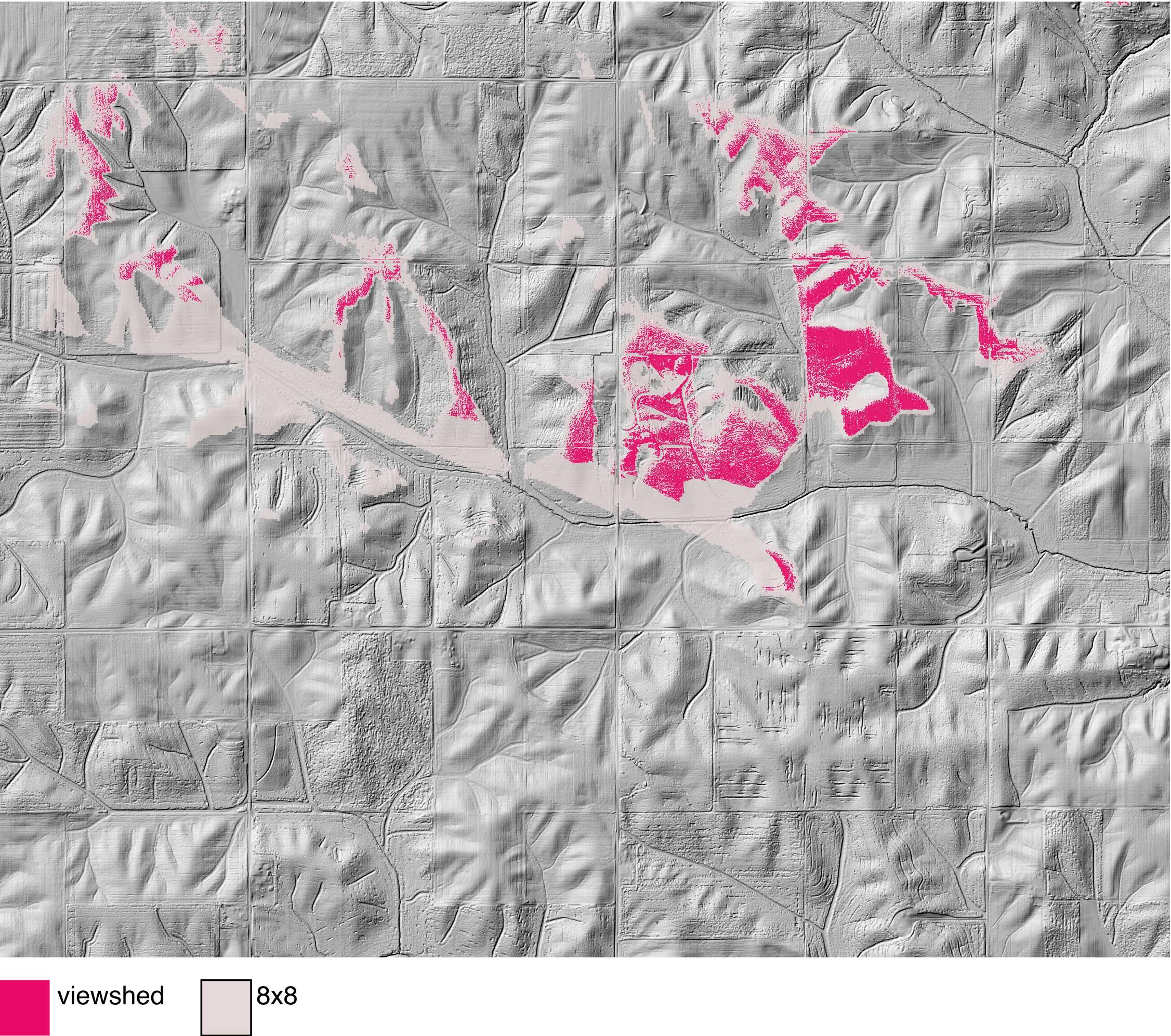}
    \caption{\lores viewshed $k=8\times8$}
  \end{subfigure}
  \begin{subfigure}{.4\textwidth}
    \centering
    \includegraphics[width=\textwidth]{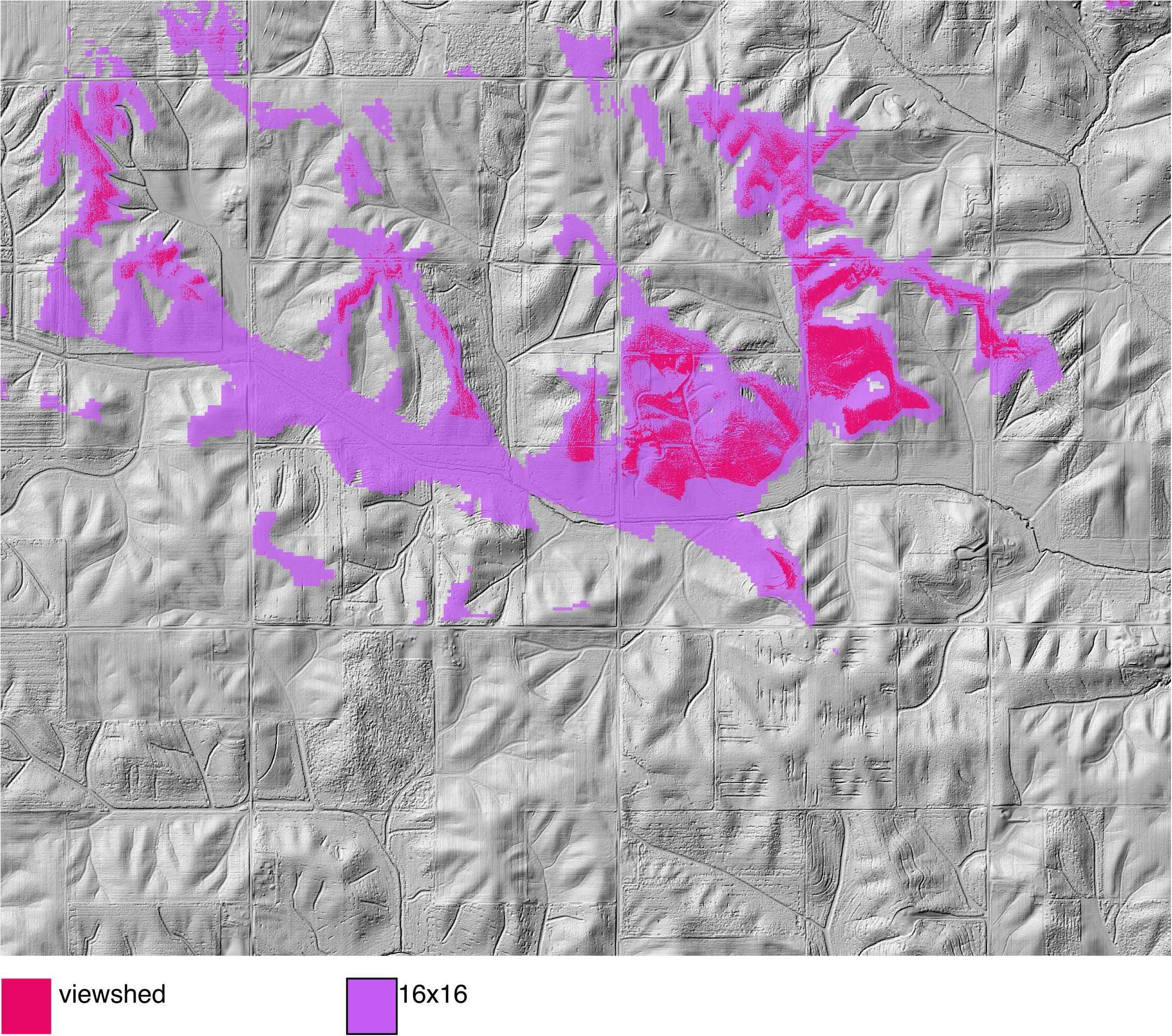}
    \caption{\lores viewshed $k=16\times16$}
  \end{subfigure}
  \begin{subfigure}{.4\textwidth}
    \centering
     \includegraphics[width=\textwidth]{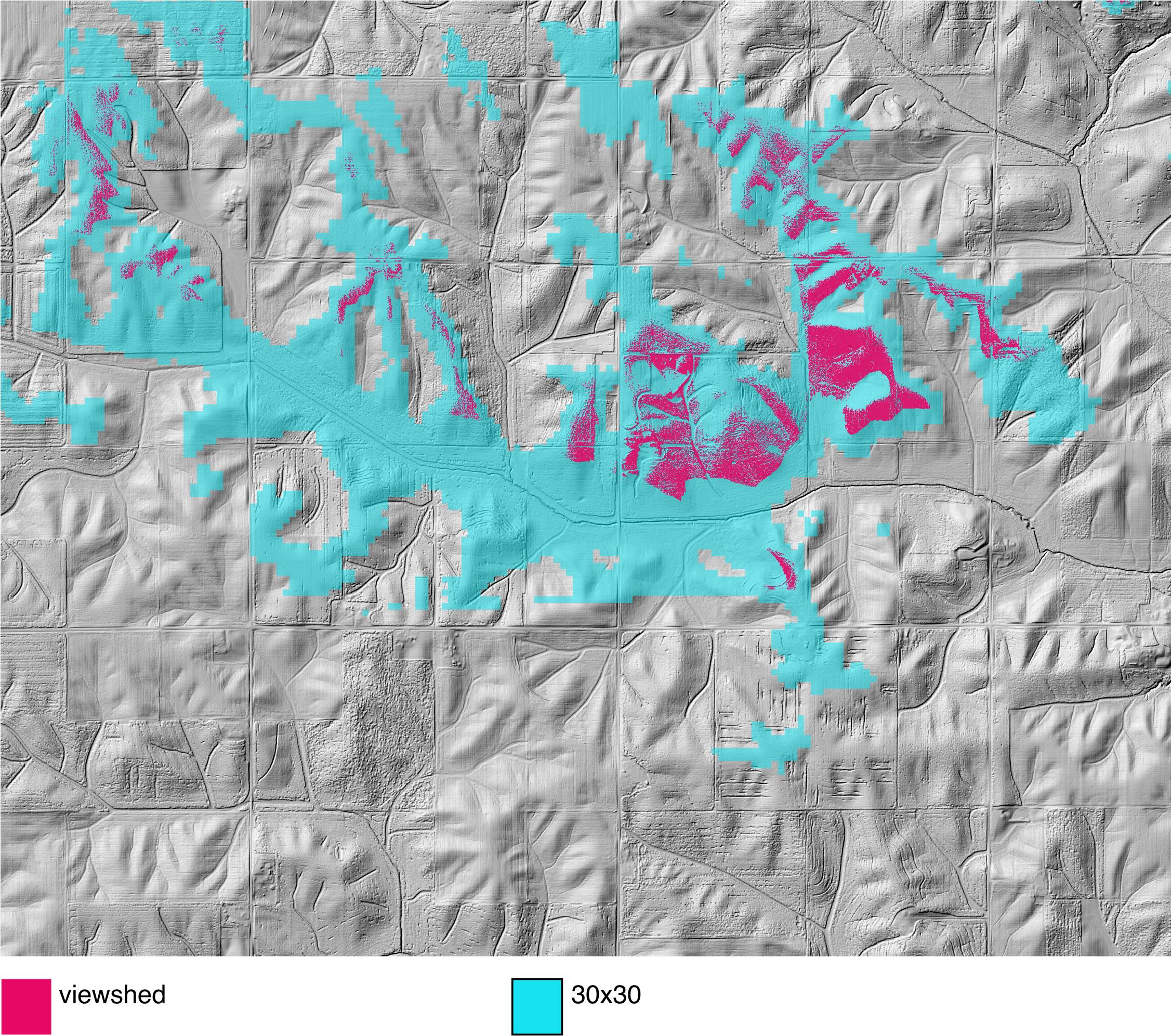}
    \caption{\lores viewshed  $k=30\times30$}
  \end{subfigure}
  \begin{subfigure}{.4\textwidth}
    \centering
      \includegraphics[width=\textwidth]{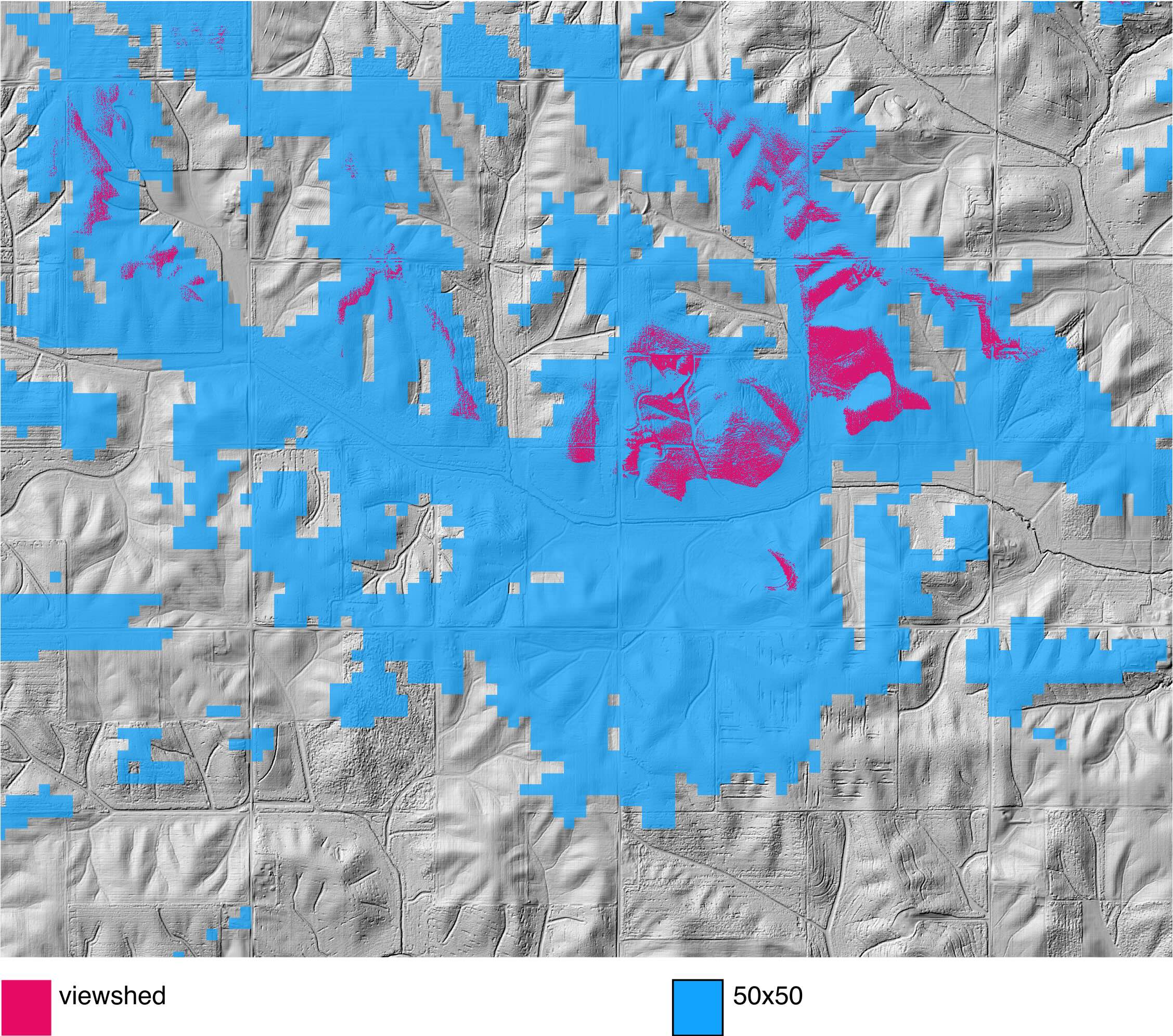}
    \caption{\lores viewshed  $k=50\times50$}
  \end{subfigure}
  \begin{subfigure}{.4\textwidth}
    \centering
     \includegraphics[width=\textwidth]{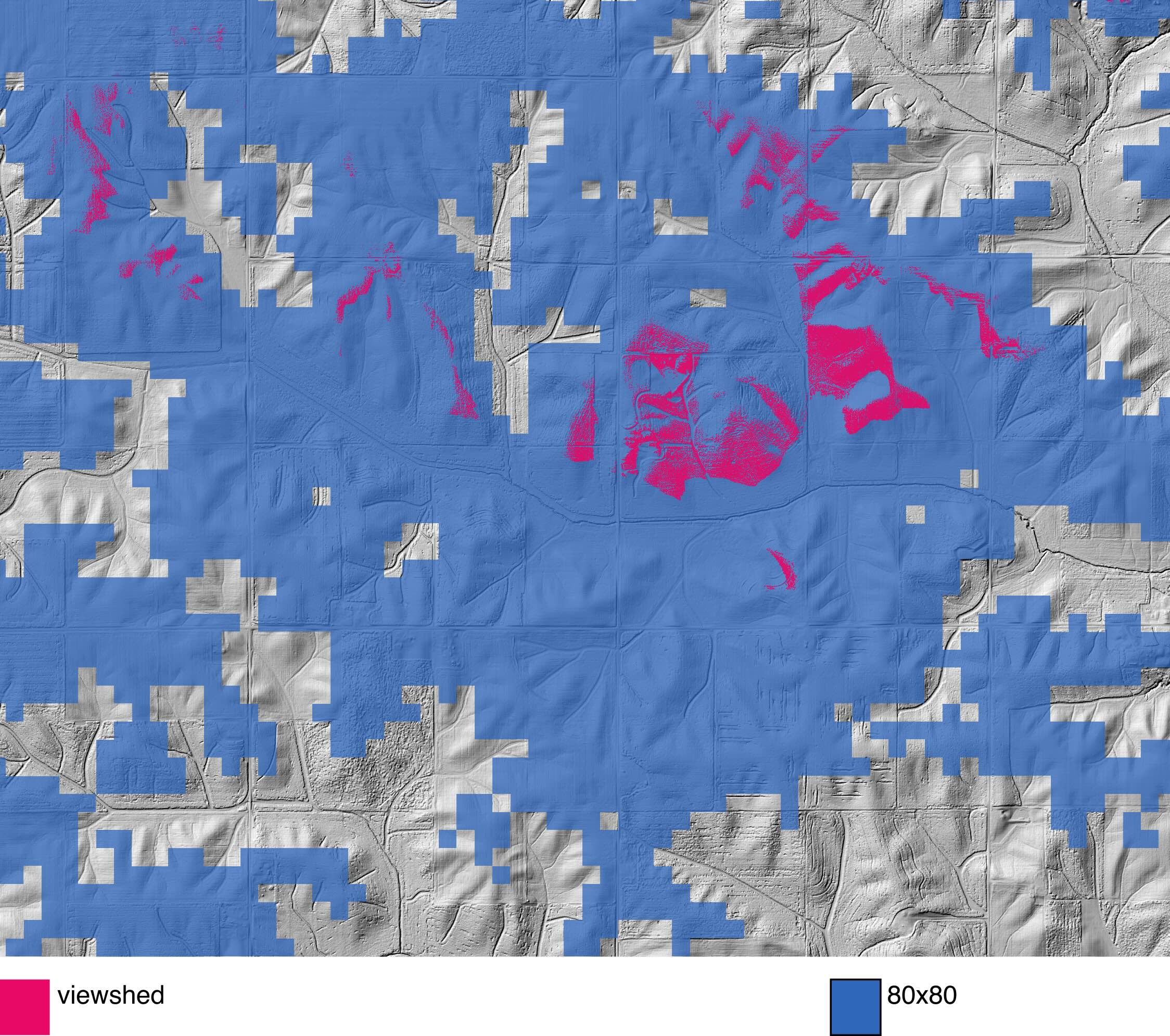}
    \caption{\lores viewshed  $k=80\times80$}
  \end{subfigure}
  \begin{subfigure}{.4\textwidth}
    \centering
      \includegraphics[width=\textwidth]{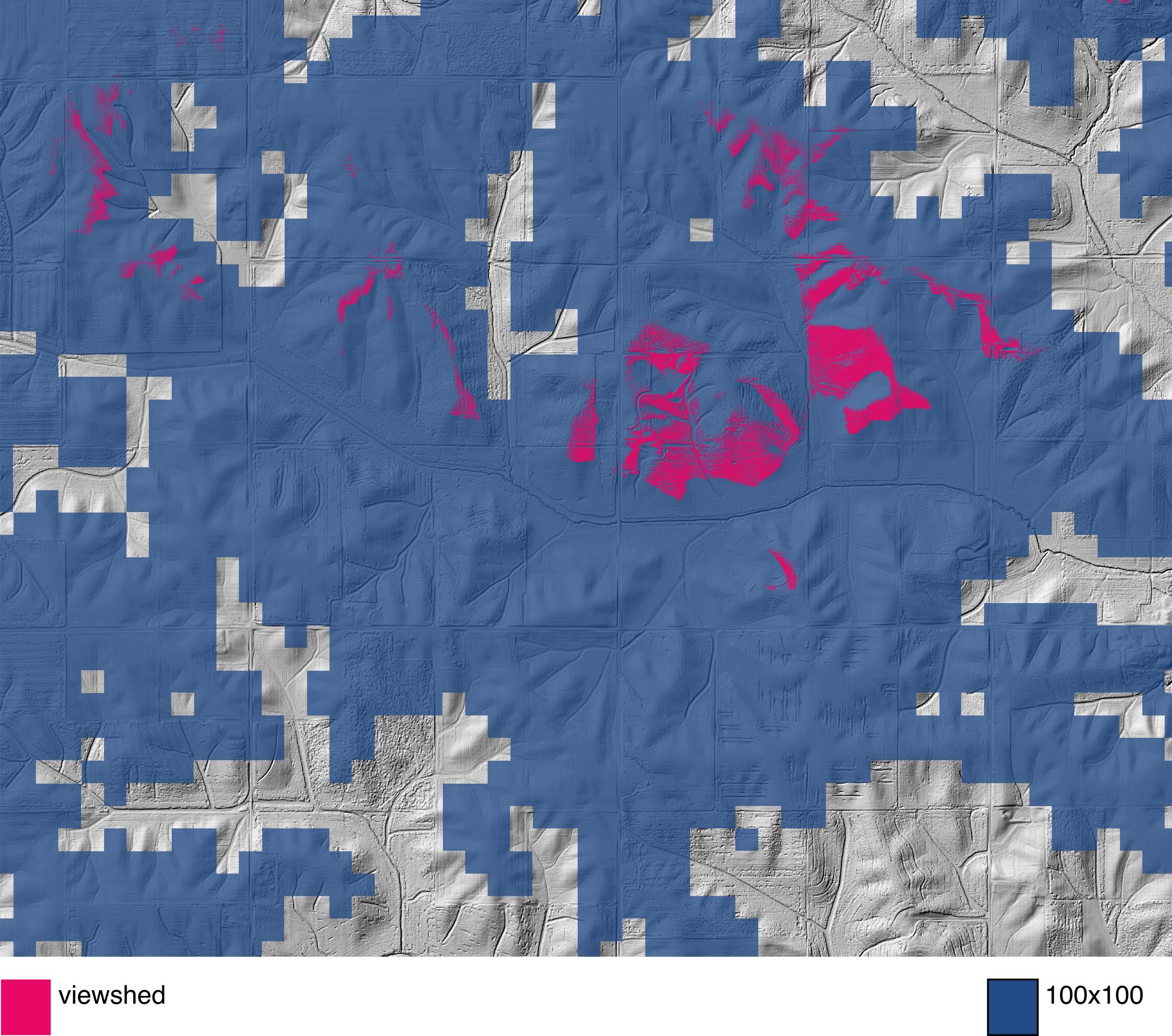}
    \caption{\lores viewshed $k=100\times 100$}
  \end{subfigure}
  \begin{subfigure}{.4\textwidth}
    \centering
      \includegraphics[width=\textwidth]{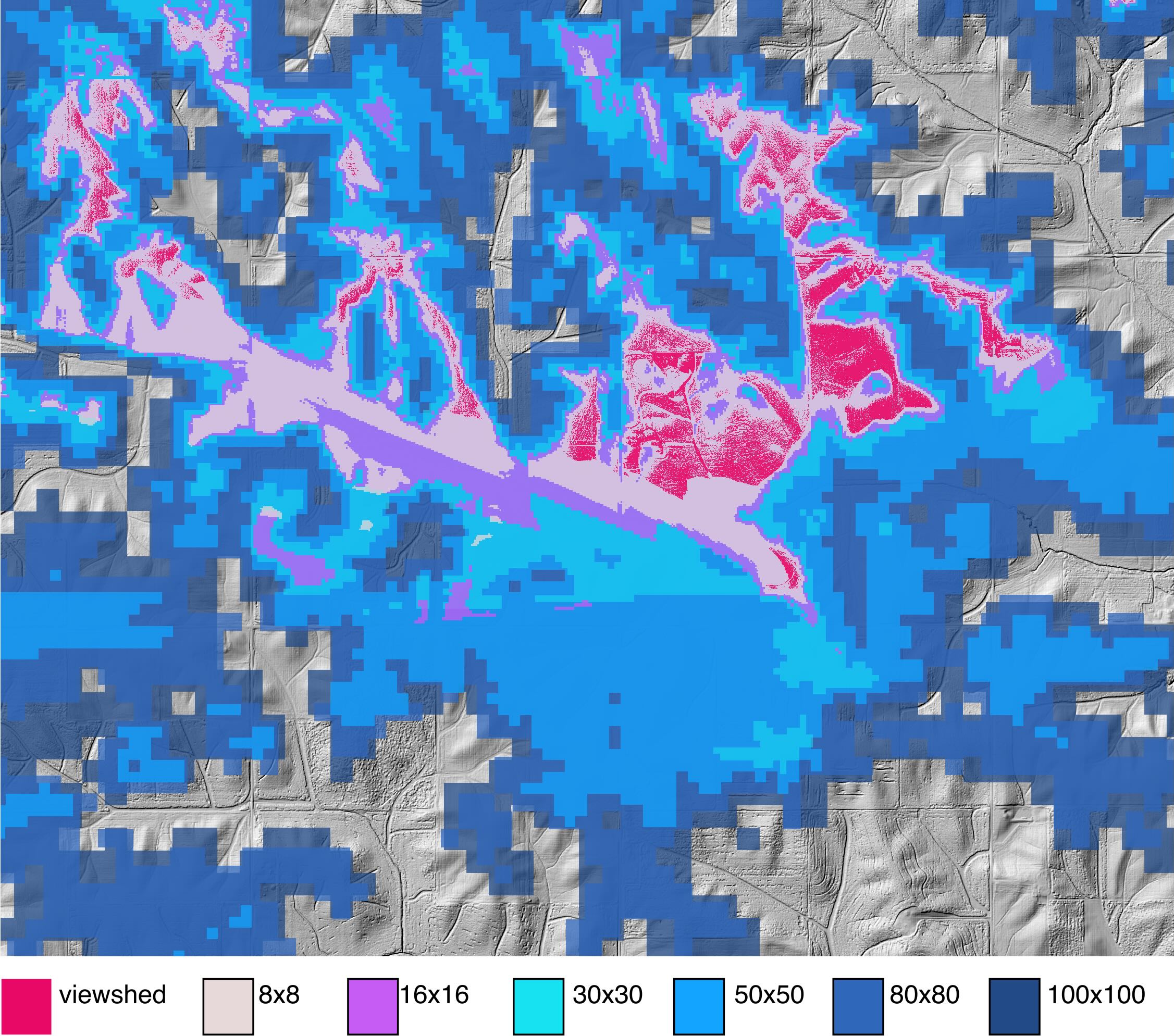}
    \caption{Overall: Effect of block size}
  \end{subfigure}
  
  \caption{\clearcr zoom in: The viewshed overlaid on the \lores
    viewsheds computed with various block
    sizes for viewpoint at     $(r=4,000, c=7,000)$.
}
\label{fig:clear_creek_zoom}
\end{figure*}

\begin{figure*}[h]
  \centering
  \begin{subfigure}{.32\textwidth}
    \centering
    \includegraphics[width=\textwidth]{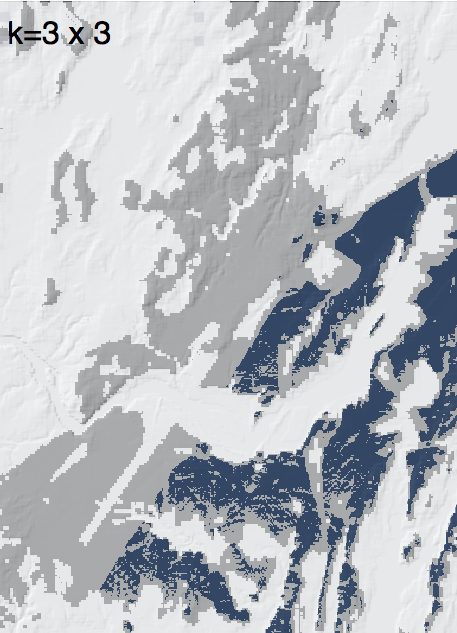}
     \end{subfigure}
  \begin{subfigure}{.32\textwidth}
    \centering
    \includegraphics[width=\textwidth]{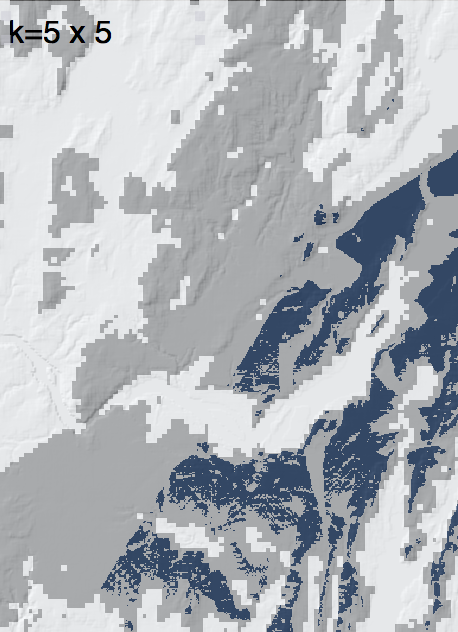}
     \end{subfigure}
    \begin{subfigure}{.32\textwidth}
  \centering
  \includegraphics[width=\textwidth]{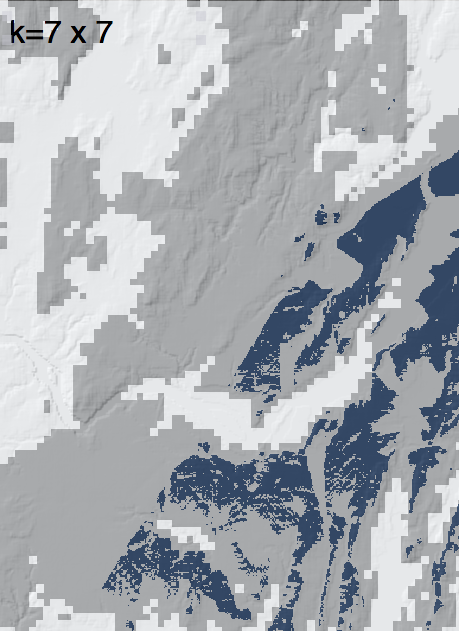}
 \end{subfigure}

\begin{subfigure}{.32\textwidth}
    \centering
    \includegraphics[width=\textwidth]{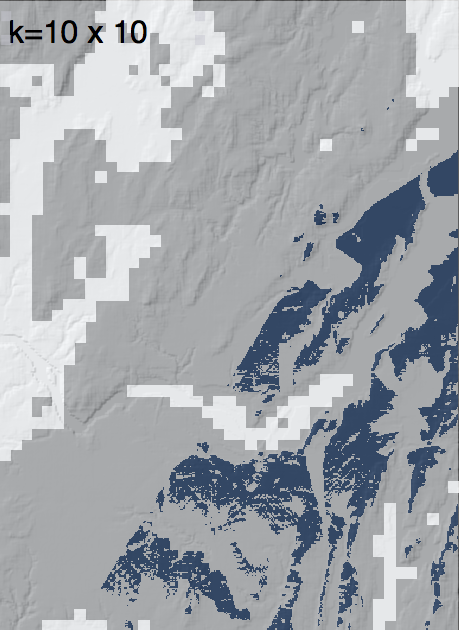}
   \end{subfigure}
  \begin{subfigure}{.32\textwidth}
    \centering
    \includegraphics[width=\textwidth]{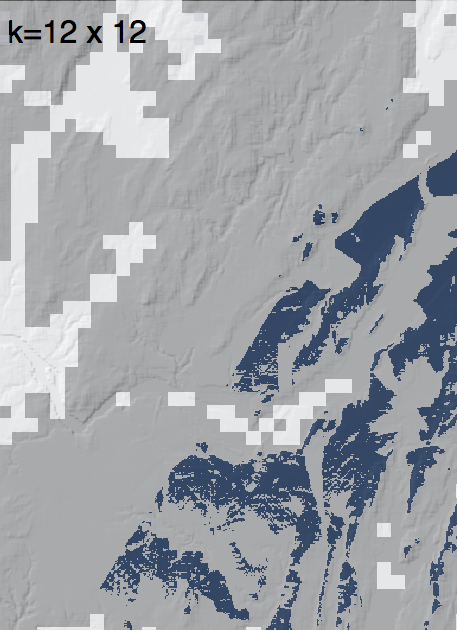}
   \end{subfigure}
    
\caption{Block size effect on \bruns dataset  for viewpoint at
  $(r=200,c=250)$. The viewshed (dark blue) and the \lores viewshed  (dark gray)
  with various block size $k$ are overlaid on the hill-shaded terrain
  (light gray). }
\label{fig:brunsdem_blocksize} \end{figure*}

\begin{table*}[h]
  \caption{\bruns ($n=205,884$) results: Running time (in seconds) for  \multires; 
    the speedup relative to \visiter for various block sizes; and
    \lores and \hires viewshed sizes (as a percentage of $n$)} 
  \label{fig:bruns_time}
  
  \begin{tabular}{@{} c c r r r  c c c c @{} }
    \toprule
    $k$ & \blockit & \lores & \hires &   \multires 
    & Speedup & \lores & \hires & \lores/\hires \\ 

  & time & time &  time &  time & &  size &  size & size \\ 

\midrule
    $3\times 3$  & 0.07      & 1,332.92 & 804.91   & 2,137.83 & 2.37   & 7.76\%       & 2.38\%       & 3.28        \\
    $4\times 4$  & 0.03      & 607.31   & 940.19   & 1,547.50 & 3.27   & 11.36\%      & 2.38\%       & 4.78   \\
    $5\times 5$  & 0.02      & 349.54   & 1,091.89 & 1,441.43 &3.51    & 15.10\%      & 2.38\%       & 6.35           \\
    $6\times 6$  & 0.02      & 230.93   & 1,163.59 & 1,394.52 & 3.63   & 18.76\%      & 2.38\%       & 7.88     \\
    $7\times 7$  & 0.01      & 163.28   & 1,358.81 & 1,522.09 &  3.33    & 22.74\%      & 2.38\%       & 9.55                       \\
    $8\times 8$  & 0.03      & 122.86   & 1,455.48 & 1,578.34 &        3.21     & 26.60\%      & 2.38\%       & 11.17     \\
    $9\times 9$  & 0.01      & 96.63    & 1,616.73 & 1,713.36 &        2.95     & 30.62\%      & 2.38\%       & 12.86   \\
    $10\times 10$ & 0.01      & 76.23    & 1,713.98 & 1,790.21 &
    2.83  & 34.04\%      & 2.38\%       & 14.30           \\ 
\bottomrule
  \end{tabular}
\end{table*}


\begin{figure*}[h]
  \begin{subfigure}{.45\textwidth}
    \includegraphics [width=.9\linewidth] {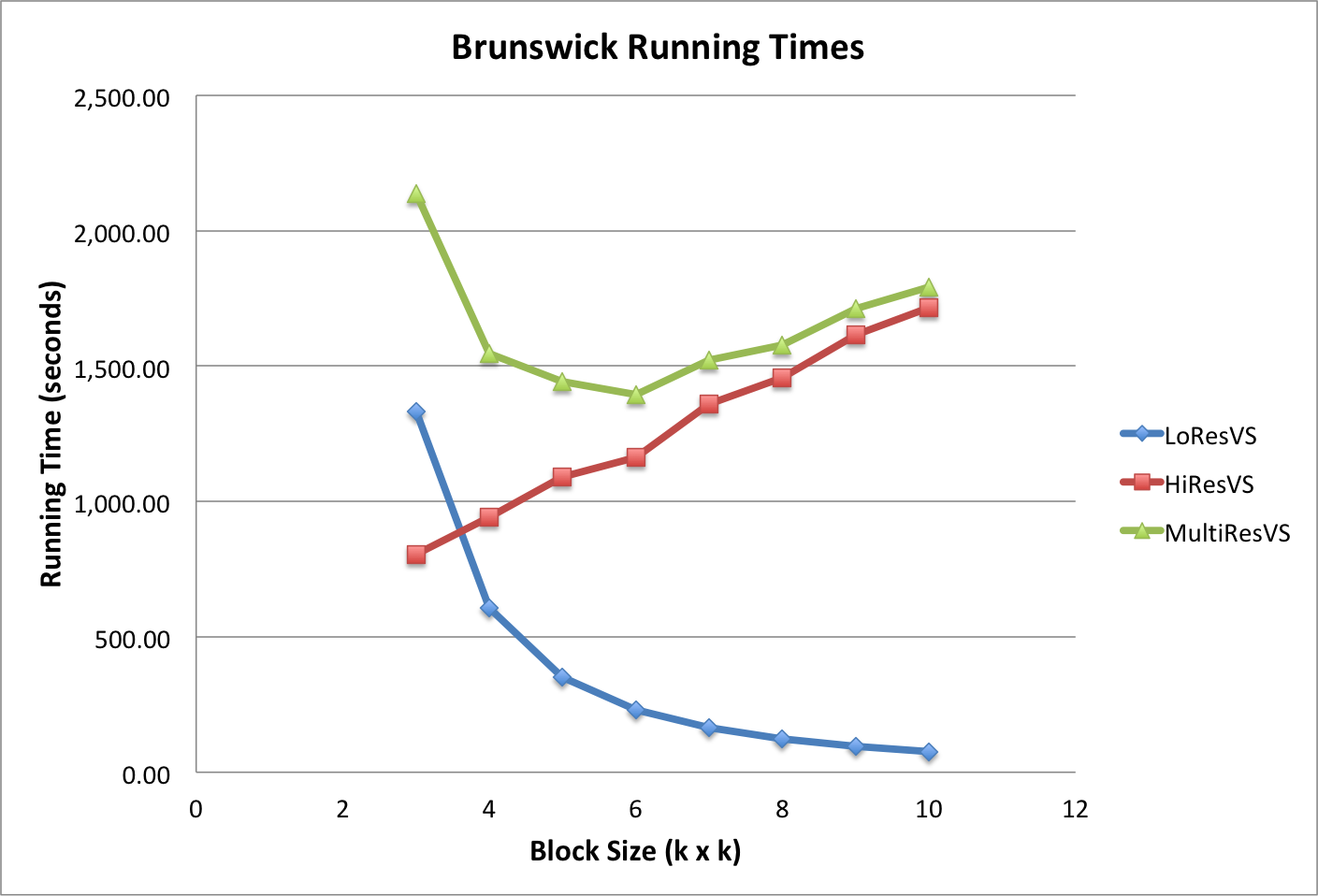}
    \caption{Running time of \multires on \bruns}
  \end{subfigure}
  \begin{subfigure}{.45\textwidth}
    \includegraphics [width=.9\linewidth] {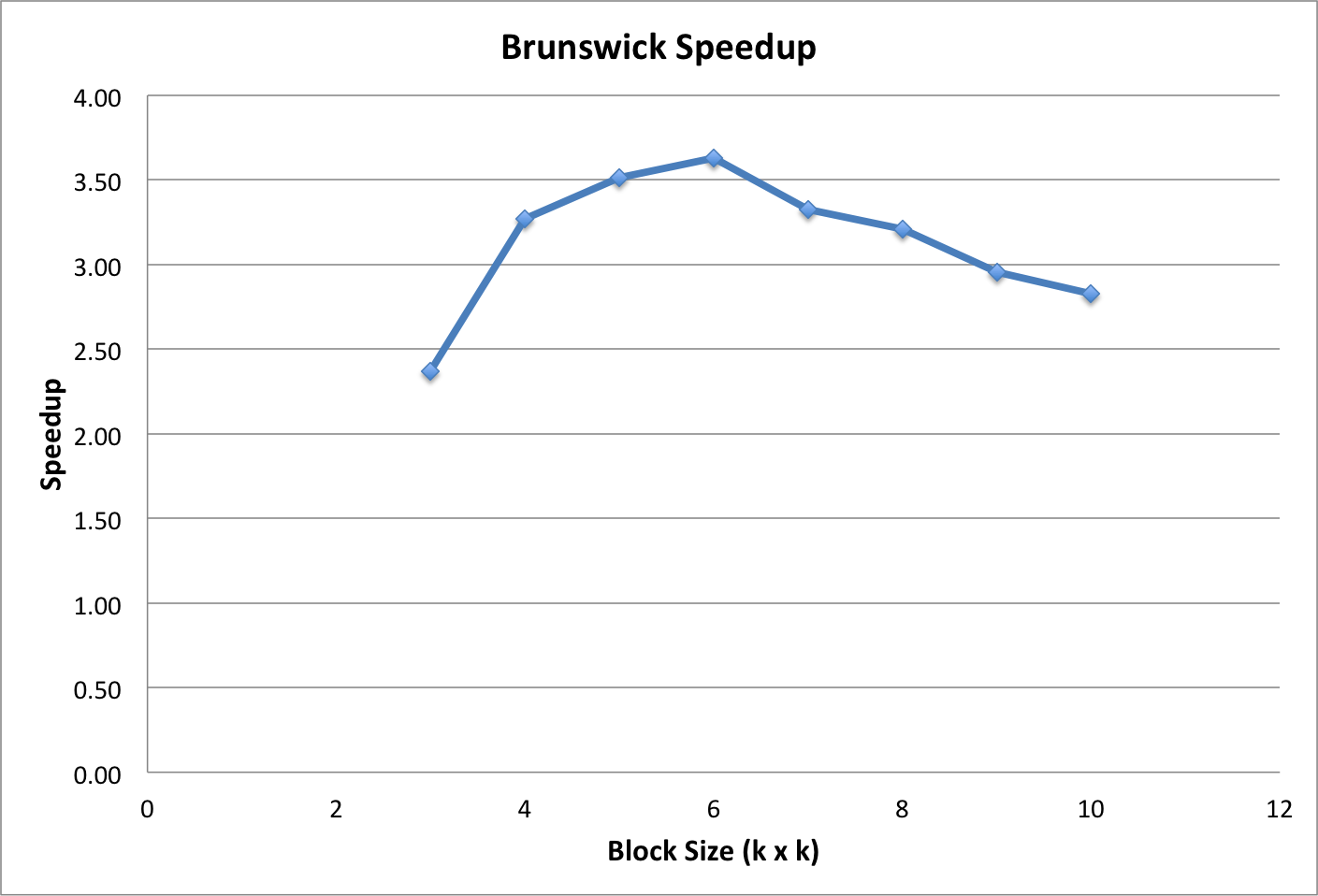}
    \caption{Speedup of \multires compared to    \visiter}
  \end{subfigure}
  \caption{Brunswick results: running times and speedup with various
    block sizes}
  \label{fig:bruns_plots}
\end{figure*}

\newpage 
\bibliographystyle{ACM-Reference-Format}
\bibliography{vis}

\end{document}